\documentclass[10pt,letterpaper]{article}

\usepackage[top=0.85in,left=2.75in,footskip=0.75in]{geometry}

\usepackage{amsmath,amssymb}

\usepackage{changepage}

\usepackage[utf8x]{inputenc}

\usepackage{textcomp,marvosym}

\usepackage{cite}

\usepackage{nameref,hyperref}

\usepackage{microtype}
\DisableLigatures[f]{encoding = *, family = * }

\usepackage[table]{xcolor}

\usepackage{xcolor,soul}

\usepackage{array}

\usepackage{eurosym}

\usepackage{todonotes}

\newcolumntype{+}{!{\vrule width 2pt}}

\newlength\savedwidth

\raggedright
\usepackage[aboveskip=1pt,labelfont=bf,labelsep=period,justification=raggedright,singlelinecheck=off]{caption}

\makeatletter
\renewcommand{\@biblabel}[1]{\quad#1.}
\makeatother

\date{}

\usepackage{lastpage,fancyhdr,graphicx}
\usepackage{epstopdf}
\fancyheadoffset[L]{2.25in}
\fancyfootoffset[L]{2.25in}
\usetikzlibrary{patterns}
\usetikzlibrary{positioning}
\usetikzlibrary{shapes}

\begin{document}
\vspace*{0.2in}

\begin{flushleft}
{\Large
\textbf\newline\ Improving Pest Monitoring Networks in order to reduce pesticide use in agriculture} 

Marie-Jos\'ee Cros\textsuperscript{1,*},
Jean-No\"el Aubertot\textsuperscript{2},
Sabrina Gaba\textsuperscript{3,4},
Xavier Reboud\textsuperscript{5},
R\'egis Sabbadin\textsuperscript{1},
Nathalie Peyrard \textsuperscript{1}
\\
\bigskip
\textbf{1} MIAT, Université de Toulouse, INRA, F-31320 Castanet-Tolosan, France
\\
\textbf{2} AGIR, Université de Toulouse, INPT, INRA, F-31320 Castanet-Tolosan, France
\\
\textbf{3} USC 1339, Centre d’Etudes Biologiques de Chizé, INRA, 79360 Villiers-en-Bois, France 
\\
\textbf{4} UMR 7372 Centre d’Études Biologiques de Chizé, CNRS \& Univ. La Rochelle, 79360 Beauvoir-sur-Niort, France
\\
\textbf{5} Agroécologie, AgroSup Dijon, INRA, Univ. Bourgogne, Univ. Bourgogne
Franche-Comté, F-21000, Dijon, France
\\
\bigskip

*Marie-Josee.Cros@inra.fr
\end{flushleft}

\section*{Abstract} 

Disease and pest control largely rely on pesticides use and  progress still remains to be made towards more sustainable practices. Pest Monitoring Networks (PMNs) can provide useful information for improving crop protection by restricting pesticide use to the situations that best require it. However, the efficacy of a PMN to control pests may depend on its spatial density and space/time sampling balance. Furthermore the best trade-off  between the monitoring effort  and the  impact of the PMN information may be pest dependent.
We developed a  generic simulation model that links  PMN information  to treatment decisions and pest dynamics. We  derived the number of treatments, the  epidemic extension and the global gross margin for different families of pests. 
For soil-borne pathogens and weeds, we found that increasing the spatial density of a PMN significantly decreased the number of treatments (up to 67\%), with an only marginal increase in infection. Considering past observations had a second-order effect (up to a 13\% decrease).
For the spatial scale of our study,  the PMN  information had practically no influence in the case of insects.
The next step is to go beyond PMN analysis to design and chose among sustainable management strategies at the landscape scale.

\section*{Keywords}
Pest monitoring system; Regional Action Plan for Pesticide Reduction; Decision rules; Oilseed rape; Dynamic Bayesian Network; Simulation

\section{Introduction}

Climate change has been identified as a major cause of emerging or re-merging diseases and animal pests in agriculture, as well as the geographical shifts of weeds. This may have an adverse impact on food security. At this time,  pest control mainly relies on pesticide use, which has been recognized for its harmful impact on biodiversity and human health. Sustainable strategies for pest management are therefore needed. One way to achieve this objective is to obtain accurate information about pest dynamics. Such information is provided by Pest Monitoring Systems (PMS) that monitor the main pests present in commercial fields. These PMS have been implemented in many countries and include the Biovigilance Network in France \cite{MAAF15}. The synthesis provided by SCAR \cite{SCAR2013} maps out the current situation on pest monitoring systems (PMS) across Europe in terms of integrated pest management aims: forecast and monitoring of pests and diseases, warning and alert systems to detect thresholds and advisory services on integrated pest management \cite{Jorgensen1996, Krocher2007, Delos2008}. Pest monitoring  systems are  not limited  to  computer-based systems to assist farmers in making the right decision but comprise initiatives, networks, activities, and tools from a wide range of  players including  farmers, advisers, monitors,  government representatives,  industry, researchers, etc. This review pinpoints the fact that “The  supervision  of  pest  monitoring  systems  and  evaluation  is  mostly  in  responsibility  of governmental institutions (Czech  Republic, Denmark, Netherlands, Poland, Sweden and Turkey) in  cooperation and support by national stakeholders such as boards  of  agriculture, national and regional advisory services and research institutions. In some countries (Belgium, Estonia, Finland, Germany and  Sweden) research institutes and advisory organizations (Ireland) are diversely fulfilling  the role of  running  statutory monitoring activities, the development of  efficient  pest and disease monitoring  tools and techniques, and  research activities related to pest monitoring systems".
Budgets are mainly allocated by national funding.”

PMNs provide information about the sanitary status of a subset of fields in the agricultural domain, which will be called upon to provide disease incidence maps \cite{Wellings2011, Jiang2017}, to build  estimators of the regional and site-specific disease incidence \cite{Metal16}, to issue plant health bulletins \cite{BSV} for stakeholders, and to shed light on the potential distribution of pests using modelling approaches \cite{Kriticos2015}, and even to carry out the retrospective reconstruction of preferential invasion  pathways \cite{Botella2018, Mack2000}.

 If the PMN provides accurate information in real time, this can help to reduce the temporal delay between the pest emergence (or re-emergence) and the warning of decision-makers who can then  make treatment decisions only when necessary, thus limiting both the immediate and future requirements for pesticide use.

Pest and disease monitoring is performed by experts who visit each field within the PMN. 
Such monitoring induces time and money costs that limit the number of sites visited, i.e., the spatial frequency and temporal extent of the networks. The spatial extent of the network is also an important feature, which will be discussed in the Results section.
It is therefore crucial that the aggregated information built from the PMN data significantly increase disease control and reduce pesticide use.
However, PMN design remains complex since there is no clear knowledge about which spatial frequency (i.e., the number of fields selected for monitoring) and temporal extent (i.e., the depth of the data history used to build the aggregated information) affect disease spread and the amount of pesticides used.\\
A wide variety of plant pests and diseases threaten crops, causing significant losses to farmers and threatening food security. These vary by their life cycle and their ecology. For instance, soil-borne plant pathogens i.e., bacteria, fungus and nematodes, spread at short distances but can survive for long periods without a host plant \cite{Hughes1996}. Since weeds are annual plants, they can also persist for decades in the seed bank and can disperse in surrounding habitats \cite{Bourgeois2019}. Finally, insect pests (as well as the viruses they harbour) have considerable spatial dispersal abilities but generally low persistence (e.g., up to 3 years for diapausing species \cite{Bardner1974}. Consequently,
the spatial frequency and extent and  the temporal extent of an efficient PMN is most likely to be pest-dependent. 
For instance, we expect that a PMN with a large spatial extent, small spatial frequency and small temporal extent to be more efficient to obtain meaningful information on species that have high spatial dispersion capabilities, whereas a PMN with a small spatial extent, large temporal extent but with high spatial frequency will be more accurate for species with low dispersal capabilities. For these reasons, and also because there is uncertainty about disease dynamics, about the way that stakeholders use the information and about the efficiency of treatment, designing an "optimal" PMN is out of reach.
It is easier to compare the efficiency of a set of different PMN candidates. However, this comparison cannot be performed \textit{in vivo} in real agricultural areas because of the human and financial costs of such experiments. 

In this study,
we developed a spatio-temporal stochastic Dynamic Bayesian Network model (DBN, \cite{JN07}) to compare PMNs with different spatial frequencies and temporal extents and we consider a fixed spatial extent of 100 ha. This model make it possible to  simulate the spatio-temporal dynamics of an epidemic in a set of crop fields for a  given pest management strategy built from the observations provided by the PMN.
To model the decision rule (treatment or not) for a given field, we make the assumption that some private information about the infection status of the field in the previous year(s) is used to modulate the public PMN information. 
The DBN model is  generic enough to model various families of pests. We use it with different sets of parameters corresponding to three  pest types that have contrasted dispersal characteristics (soil-borne pathogens, weeds and insects). Using oil seed rape as a case study, we evaluate the epidemic size, the number of treatments and the expected gross margin  for these three pest types and for PMNs with varying spatial and temporal densities. 

For soil-borne pathogens and weeds, we found that increasing the spatial density of a PMN significantly decreased the number of treatments (up to 67\%), with only a  marginal increase in infection. Considering past observations only had a second-order effect (up to a 13\%  decrease).
For the spatial scale of our study, the PMN information had practically no influence
in the case of insects.

\section{Materials and methods}
\subsection{The model}
\subsubsection*{Dynamic Bayesian Network model of pest dynamics}
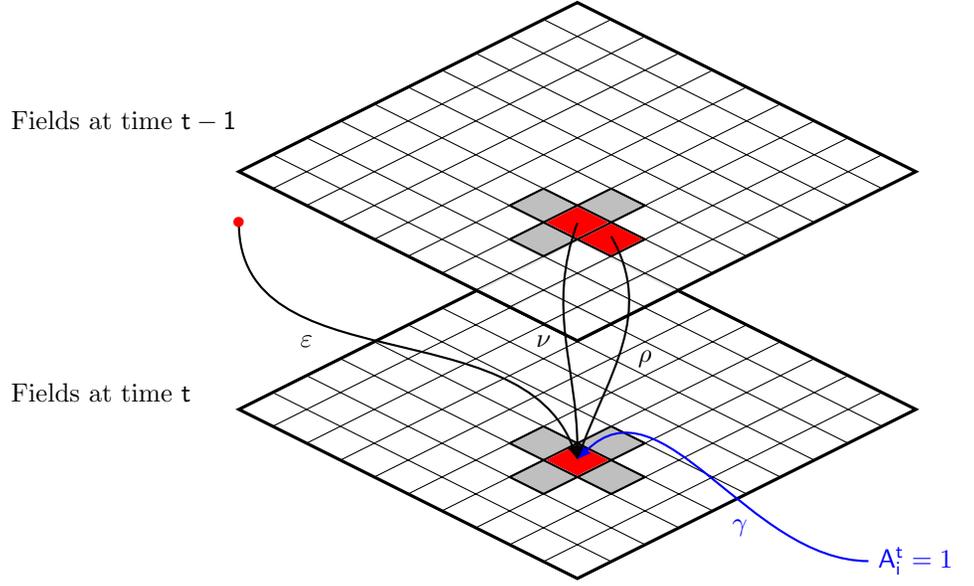
\begin{figure}
\begin{tikzpicture}[scale=.9,every node/.style={minimum size=1cm},on grid]
		
    \begin{scope}[
            yshift=-50,every node/.append style={
            yslant=0.5,xslant=-1},yslant=0.5,xslant=-1
            ]
        \fill[white,fill opacity=0.9] (0,0) rectangle (5,5);
        \draw[step=5mm, black] (0,0) grid (5,5); 
        \draw[black,very thick] (0,0) rectangle (5,5);

        \draw[black,thick,fill opacity=0.5, fill=gray] (1,1.5) rectangle (1.5,2);
        \draw[black,thick,fill opacity=0.5, fill=gray] (1.5,1) rectangle (2,1.5);
        \draw[black,thick,fill opacity=0.5, fill=gray] (2,1.5) rectangle (2.5,2);
        \draw[black,thick,fill opacity=0.5, fill=gray] (1.5,2) rectangle (2,2.5);

        \fill[red] (1.52,1.52) rectangle (1.98,1.98);
    \end{scope}
    	
    \begin{scope}[
    	yshift=50,every node/.append style={
    	    yslant=0.5,xslant=-1},yslant=0.5,xslant=-1
    	             ]
        \fill[white,fill opacity=.9] (0,0) rectangle (5,5);
        \draw[black,very thick] (0,0) rectangle (5,5);
        \draw[step=5mm, black] (0,0) grid (5,5);
        
        \draw[black,thick,fill opacity=0.5, fill=gray] (1,1.5) rectangle (1.5,2);
        \draw[black,thick,fill opacity=0.5, fill=gray] (1.5,1) rectangle (2,1.5);
        \draw[black,thick,fill opacity=0.5, fill=gray] (2,1.5) rectangle (2.5,2);
        \draw[black,thick,fill opacity=0.5, fill=gray] (1.5,2) rectangle (2,2.5);
        
        \fill[red] (1.52,1.52) rectangle (1.98,1.98);
        \fill[red] (1.52,1.02) rectangle (1.98,1.48);
        
    \end{scope}
        
        \draw (-8.5,1) node[right]{Fields at time $\mathsf{t}$};
        \draw (-8.5,5) node[right]{Fields at time $\mathsf{t-1}$};


    \draw[-latex,thick,black](0,3.5)
        to[out=250,in=90] (0,0);
    \node at(-0.5,1.75) {$\mathbf{\nu}$};
    
    \draw[-latex,thick,black](0.5,3.3)
        to[out=300,in=70] (0,0);
    \node at(1,1.5) {$\mathbf{\rho}$};
    
    \draw[-latex,thick,black](-5,3.5)
        to[out=270,in=110] (0,0);
    \node at(-4,1.75) {$\mathbf{\varepsilon}$};
    \node[red] at(-5,3.5) {$\bullet$};

    \draw[-latex,thick,blue](4.3,-1.5)node[right]{$\mathsf{A_i^t}=1$}
        to[out=180,in=45] (0,0);
    \node[blue] at(2.4,-1) {$\mathbf{\gamma}$};
\end{tikzpicture}
\caption{{\bf Pest dynamic parameters of a field $i$.} $\varepsilon$, long-distance dispersal probability; $\rho$, short-distance dispersal probability; $\nu$, infection persistence probability ; 
($\gamma$) treatment efficiency.
Red squares: infected fields ; gray squares: neighbours of field $i$.
\label{fig:Dynamics}}
\end{figure}

We modelled the spatio-temporal dynamics of a disease in a set of crop fields within the framework of a  DBN. A DBN is a particular case of a Bayesian Network where variables are indexed by time, and the state of variables at time $t$ depends  on the state of variables at time $t-1$ (Markovian assumption).
The state of a field $i$ at time $t$ is a random variable denoted $X_i^t$, and can take two values: 0  for a non-infected field, and 1 for a field where the pest is present. This state may depend on the state of every  field in the neighbourhood of the field $i$ at time $t-1$, denoted  $X^{t-1}_{N_i}$ (similarly to  the spatial SIS model \cite{Harris74}) and on the action $A^t_i$ applied at time $t$. We consider two actions: $A^t_i = 1$ if the treatment  is applied and $A^t_i = 0$ otherwise.  The probabilities of transition from  state $X^{t-1}_i$ to state $X^{t}_i$, given the action applied $A^t_i$ and the state of the neighbourhood $X^{t-1}_{N_i}$ are parameterized by $\epsilon$, the long-distance dispersal probability of the pest,
$\rho$  the probability of infection from a neighbouring infected field, 
$\nu$  the probability of pest survival between $t-1$ and $t$ if no treatment is applied and
$\gamma$ the probability of pest eradication after treatment (see figure{~\ref{fig:Dynamics}}). The probabilities of pest arrival with and without treatment are
\begin{align*}
P(X^{t}_i = 1 \mid X^{t-1}_i = 0, X^{t-1}_{N_i}, A^t_i = 0) &=  \epsilon + (1 -\epsilon)(1 - (1-\rho)^{K^{t-1}_i}) = P_{01}(K^{t-1}_i),\\
P(X^{t}_i = 1 \mid X^{t-1}_i = 0, X^{t-1}_{N_i}, A^t_i = 1) &= (1-\gamma)P_{01}(K^{t-1}_i),\\
\end{align*}
where $K_i^{t-1}$ is the number of infected neighbours of site $i$ at time $t-1$.

\noindent The probabilities of persistence  with and without treatment are:
\begin{align*}
P(X^{t}_i = 1 \mid X^{t-1}_i = 1, X^{t-1}_{N_i}, A^t_i = 0) &=  \nu + (1-\nu)P_{01}(K^{t-1}_i) = P_{11}(K^{t-1}_i),\\
P(X^{t}_i = 1 \mid X^{t-1}_i = 1, X^{t-1}_{N_i}, A^t_i = 1) &= (1-\gamma)P_{11}(K^{t-1}_i).\\
\end{align*}
At the landscape level, the  global probability of transition is:
$$P(X^{t}|X^{t-1}, A^t) = \prod_{i=1}^n P(X_i^{t}|X_i^{t-1}, X_{N_i}^{t-1}, A_i^t),$$
where $X^t = (X^t_1, \ldots, X^t_n)$ and $A^t= (A^t_1, \ldots, A^t_n)$ are, respectively the vectors of sanitary states of and actions applied to every field at time $t$. For a given number $n$ of fields and their neighbourhoods, and a given sequence of actions, the DBN model depends on four parameters that are summarized  in Table{~\ref{tab:Variables}} (top).
Action $A^t_i$ is the result of a decision rule applied at field $i$, and which depends of the information provided by the PMN, the private information of the sanitary status of field $i$, and economic parameters. We describe these different elements below  and  in Table{~\ref{tab:Variables}} (bottom).

\begin{table}[!ht]
\begin{tabular}{c l}
\cellcolor{lightgray}  Parameter  &  \cellcolor{lightgray} Definition  \\
\multicolumn{2}{l}{\textbf{Pest dynamics parameters}} \\
$\epsilon$ & long-distance dispersal probability \\
$\rho$   &  probability of infection from a neighbouring infected field \\
$\nu$  & probability of pest survival if not treated  \\
$\gamma$ & probability of treatment efficacy\\
\multicolumn{2}{l}{\textbf{Economic parameters} }\\
$y$    & maximal annual yield, in kg/ha      \\
$q $     & proportion of remaining yield when the field is infected\\
$price$    & selling price, in {\euro}/kg     \\
$c$        & all production costs (seeds, fertilizer, labour,  
            pesticide) in {\euro}/ha\\
$c_{pest}$       & annual cost of  treatments for a given pest, in {\euro}/ha    \\
\hline \\
\end{tabular}
\caption{\bf{Parameters of the DBN model for pest dynamics and the decision rule for treatment.}}
\label{tab:Variables}
\end{table}

\subsubsection*{PMN information}
A PMN is a subset $O \subseteq \{1, \ldots, n\}$ of the  $n$ fields, which  are monitored at each time step. The state $X_o^{t-1}$ for $o \in O$ is therefore  public information available to all farmers at time $t$. 
The spatial density of a PMN is the number of fields in $O$.
The temporal density is  the number of past years (or history depth) of observations considered to build the aggregated indicators that will be used in the decision rule.
If the temporal density is $h$, then the decision to treat field $i$ at time $t$ will be  based on the knowledge of all $X_o^{t'}$ for the previous $h$ time steps: $t' \in \{t-h,t-h +1,\ldots,t-1\}$ (less if $t<h$).
The whole set of PMN observations, in space and time, are then aggregated into indicators $\{p_i^t\}_{i=1,..,n}\in [0,1]$, where $p_i^t$ is the marginal probability of infection of site $i$ at time $t$ in the DBN model, conditionally to the PMN observations (see figure{~\ref{fig:PMNmodel}}).

\subsubsection*{Pest management strategy}
We considered a pest management strategy implemented yearly at the field level (the time step of the DBN is therefore the year).
The decision $A_i^t$  applied to field $i$  at time $t$ is the result of a  decision rule (referred to as $d$) that
aims at maximizing the expected gross margin $m_i^t$ of the field at time $t$. This margin is the difference between the income and the costs. The income is the product of yield price ($price$) and  yield. The yield itself depends on the maximal annual yield ($y$), the probability of infection ($p_i^t$),  the proportion of remaining yield when infected  ($q$), and the probability of treatment  efficiency ($\gamma$) (see Table \ref{tab:Variables}). The costs include or not the cost of treating the targeted pest: the production cost $c$ includes the treatment costs for all of the main pests of the crop) and,  in particular, the cost for treatment of the targeted pest, $c_{pest}$. 

If field $i$ is not treated ($A_i^t = 0$): 
$$
m_i^t(0) =  \Big((1 - p_i^t) y + p_i^t qy\Big)price - (c - c_{pest}).
$$ 
Indeed, we get the maximal yield $y$ if the field is not infected (which occurs with probability $1-p_i^t$) and only yield $qy$ when it is infected. The yield is multiplied by the price and production costs are subtracted to compute the margin. When the field is not treated, treatment cost $c_{pest}$ is subtracted from production costs $c$.

If field $i$ is treated ($A_i^t = 1$):
$$
m_i^t(1) =  \Big((1 - p_i^t) y + p_i^t \big( (1-\gamma)qy + \gamma y\big)\Big)price - c.
$$
If the treatment is efficient (probability $\gamma$), the maximal yield $y$ is obtained, whereas if the treatment does not work, only a fraction $qy$ of the maximal yield is obtained.

\begin{figure}
\begin{tikzpicture}[scale=.9,every node/.style={minimum size=1cm},on grid]
    
    \begin{scope}[
    	yshift=-150,every node/.append style={
    	    yslant=0.5,xslant=-1},yslant=0.5,xslant=-1
    	             ]
        \fill[white,fill opacity=.9] (0,0) rectangle (5,5);
        \draw[black,very thick] (0,0) rectangle (5,5);
        \draw[step=5mm, black] (0,0) grid (5,5);
           
        \fill[red!33] (1.52,1.52) rectangle (1.98,1.98);
    \end{scope}
    		
    \begin{scope}[
            yshift=-50,every node/.append style={
            yslant=0.5,xslant=-1},yslant=0.5,xslant=-1
            ]
        \fill[white,fill opacity=0.9] (0,0) rectangle (5,5);
        \draw[step=5mm, black] (0,0) grid (5,5); 
        \draw[black,very thick] (0,0) rectangle (5,5);
        \fill[red] (1.52,1.52) rectangle (1.98,1.98);
        
        \draw[black,very thick,pattern=north west lines, fill=red] (4,1) rectangle (4.5,1.5);
        \draw[black,very thick,pattern=north west lines,fill=red] (1,4) rectangle (1.5,4.5);
        \draw[black,very thick,pattern=north west lines, fill=red] (2.5,2.5) rectangle (3,3);
        \draw[black,very thick,pattern=north west lines, pattern color=gray] (1,1) rectangle (1.5,1.5);
        \draw[black,very thick,pattern=north west lines, pattern color=gray] (1,2.5) rectangle (1.5,3);
        \draw[black,very thick,pattern=north west lines, pattern color=gray] (1,4) rectangle (1.5,4.5);
        \draw[black,very thick,pattern=north west lines, pattern color=gray] (2.5,1) rectangle (3,1.5);
        \draw[black,very thick,pattern=north west lines, pattern color=gray] (2.5,2.5) rectangle (3,3);
        \draw[black,very thick,pattern=north west lines, pattern color=gray] (2.5,4) rectangle (3,4.5);
        \draw[black,very thick,pattern=north west lines, pattern color=gray] (4,2.5) rectangle (4.5,3);
        \draw[black,very thick,pattern=north west lines, pattern color=gray] (4,4) rectangle (4.5,4.5);
    \end{scope}
    	
    \begin{scope}[
    	yshift=50,every node/.append style={
    	    yslant=0.5,xslant=-1},yslant=0.5,xslant=-1
    	             ]
        \fill[white,fill opacity=.9] (0,0) rectangle (5,5);
        \draw[black,very thick] (0,0) rectangle (5,5);
        \draw[step=5mm, black] (0,0) grid (5,5);
        
        \draw[black,very thick,pattern=north west lines,fill=red] (1,1) rectangle (1.5,1.5);
        \draw[black,very thick,pattern=north west lines,fill=red] (1,4) rectangle (1.5,4.5);
        \draw[black,very thick,pattern=north west lines,fill=red] (2.5,1) rectangle (3,1.5);
        \draw[black,very thick,pattern=north west lines, pattern color=gray] (1,1) rectangle (1.5,1.5);
        \draw[black,very thick,pattern=north west lines, pattern color=gray] (1,2.5) rectangle (1.5,3);
        \draw[black,very thick,pattern=north west lines, pattern color=gray] (1,4) rectangle (1.5,4.5);
        \draw[black,very thick,pattern=north west lines, pattern color=gray] (2.5,1) rectangle (3,1.5);
        \draw[black,very thick,pattern=north west lines, pattern color=gray] (2.5,2.5) rectangle (3,3);
        \draw[black,very thick,pattern=north west lines, pattern color=gray] (2.5,4) rectangle (3,4.5);
        \draw[black,very thick,pattern=north west lines, pattern color=gray] (4,1) rectangle (4.5,1.5);
        \draw[black,very thick,pattern=north west lines, pattern color=gray] (4,2.5) rectangle (4.5,3);
        \draw[black,very thick,pattern=north west lines, pattern color=gray] (4,4) rectangle (4.5,4.5);
    
    \end{scope}
    	
    \draw (-8.5,-2.6) node[right]{Fields at time $\mathsf{t}$};

        
    \draw (-8.5,1) node[right]{Fields at time $\mathsf{t-1}$};

    \draw (-8.5,4.5) node[right]{Fields at time $\mathsf{t-h}$};
    \node at (-7,2.65) {$\mathbf{\bullet}$};
    \node at (-7,2.90) {$\mathbf{\bullet}$};
    \node at (-7,2.40) {$\mathbf{\bullet}$};
    
    \draw[blue,very thick,dashed,rounded corners] (-5.5,-2) rectangle (5.5,7);
    \node[blue] at (-7,6.5) {\Large\bf PMN};
    \draw[->,blue,very thick] (0,-2) -- (0,-3.2);
    
    \node[blue] at (0,-3.5){$p_i^t$};
    \node at (4,-5){\textcolor{blue}{$p_i^t$} $>$ \textcolor{green}{$p_0$} $\Rightarrow$ \textcolor{red}{$A_i^t=1$}};
    \draw[->,green,very thick] (0,0) to[out=0,in=45] (3.5,-4.8);
        
\end{tikzpicture}
\caption{
{\bf Treatment decision rule} $d$: based on PMN information, the probability of field $i$ being infected ($p_i^t$) is computed (blue arrow). Based on private information, the threshold ($p_0$) is set (green arrow). If the probability of infection exceeds the threshold, then treatment is applied ($A_i^t=1$).
 Dashed cells are monitored by the PMN; red ones are infected.}
\label{fig:PMNmodel}
\end{figure}
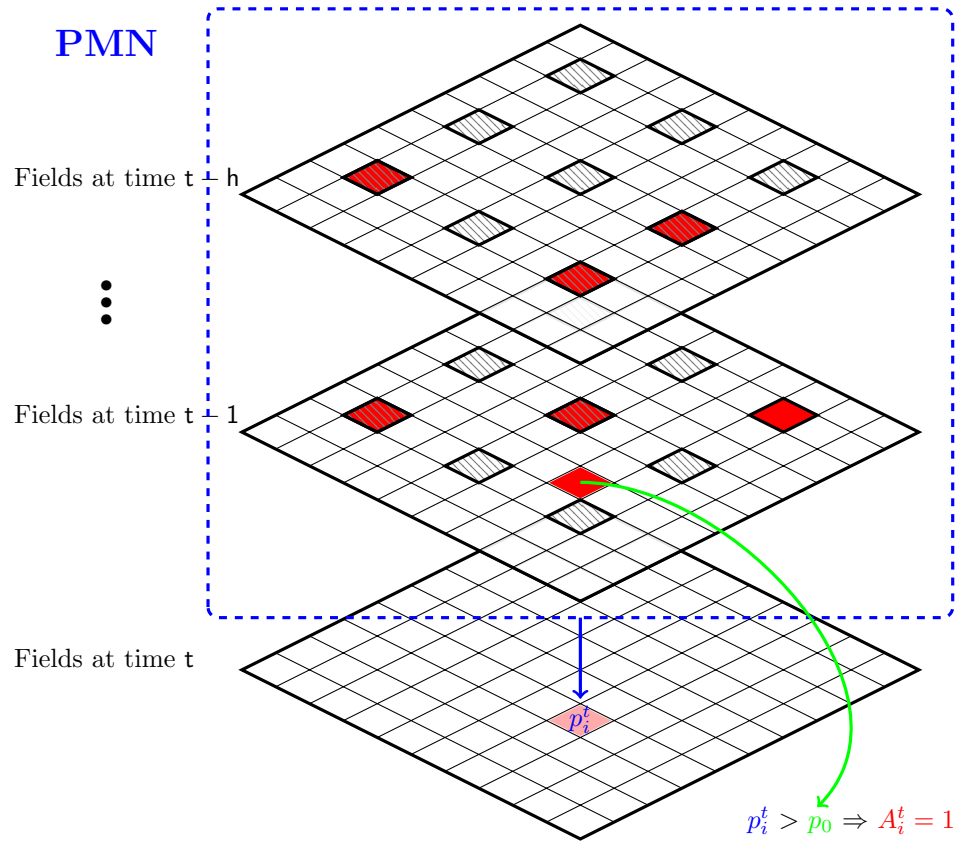

The decision rule $d$ is then to apply treatment ($A_i^t = 1$) only when $m_i^t(1) > m_i^t(0)$.
This translates into a decision to treat based on a threshold of the estimated probability of infection (see figure \ref{fig:PMNmodel}):
Treat only when
$$
p_i^t > p_0=\frac{c_{pest}}{price\gamma(1-q)y}.
$$
We then make the assumption that farmers will decide to treat a field considering not only the common PMN information but also some  private information about its history of infection. By private information, we mean that this information is only available to make a decision on that field. It is assumed to be unavailable to make a decision about other fields, even if they belong to the same farmer.
We also assume that the probability to treat will be higher in a field previously infected. We modelled this by taking  the
private information available about a field into account after the evaluation of $p_0$, by decreasing the treatment threshold when the field was  previously infected.
Specifically, considering the private information history of $h$ (i.e., the $h$ past years are considered to modulate $p_0$), we choose
\begin{itemize}
    \item If $X^{t-1}_i=1$, then $p_0$ will be decreased (treatment will be privileged). More specifically, $p_0\leftarrow p_0-\frac{k}{2^h}$, where $k$ is the number of years when field $i$ was infected in the past $h$ years.
    \item If $X^{t-1}_i=0$, the $p_0$ will be increased (treatment will less often be applied). More specifically, $p_0\leftarrow p_0+\frac{h-k}{2^h}$, where $k$ is the number of years when field $i$ was infected in the past $h$ years.
\end{itemize}

In other terms, when provided the same (public information-based) probability of infection, a farmer will be more prone to use treatment on a field when his/her field was recently infected.

\subsection{Study}
\subsubsection*{Pest types}
Three types of crop pests were considered: soil-borne pathogens, weeds and insects.
These types have contrasted spatial dispersion and temporal soil persistence: soil-borne pathogens have  low  dispersal probability (both at long  and short distances) but high persistence, weeds have  intermediate dispersal probability and high persistence, and, finally, insects have  larger dispersal probabilities but lower persistence. These different characteristics require different pesticide treatments. 
We considered the case of oilseed rape crops in France, and parameter values  were established by expertise
and studies made in France ~\cite{FOP2015, Inosys2015,Cros2017} (see Table~\ref{tab:PestParameters} in Appendix A).

\subsubsection*{Simulation protocol}
We considered $n=144$ crop fields organized on a regular grid of 12 x 12 fields, each with an area of 1 ha. The neighbourhood of a field is composed of the four fields surrounding it. Treatments and observations are only possible in the 100 fields forming the inner 10 x 10 grid  (to avoid border effects, the outer fields are not managed and form a natural reservoir for the pests). 
We compared four spatial densities for $O$: 1\%, 10\%, 25\%, 50\% of all fields, uniformly spatially distributed and stable over time, which are referred to as PMN1 to PMN4 (see figure 1 in Supplementary Information). 
Then, for each spatial density, we considered two temporal densities: the decision to treat field $i$ at time $t$ was based  either on the knowledge of $X_o^{t-1}$ only (history depth $h=1$), or on the knowledge of all PMN observations since the beginning of the simulation.

For the purpose of comparison, we simulated two extreme decision rules that do not use the PMN information: rule $d_{never}$ where fields are never treated, and rule $d_{conv}$ corresponding to a conventional management system where the standard treatments are always applied.  This results in the comparison of 10 scenarios. 

For each scenario, we considered  that  simulations start with four grouped infected fields, and we used three different positions of these first infected fields (corner, border or centre of the grid; see figure 2 in SI).
For each position of first infected fields, we generated 20 simulations of 8 time steps (8 years) that accounts for the biology of the pest. 
We considered short trajectories in order to study the short- to mid-term effect of the PMN. We ran a total of 60 simulations per scenarios (due to the computational time of a single simulation, we had to limit this number). 
Indeed, values $p_i^t$ were computed for every $i$ and $t$ using the Gibbs Sampling algorithm, which is known to be time consuming (about 6 mn per simulation on a server with Intel Xeon ES processors). We used the version implemented in the Matlab {\em Bayesian Network Toolbox} \cite{Murphy2007} with 200,000 samples.

\subsubsection*{Criteria to compare PMN}
In order to study the influence of the information provided by PMN about different spatial and temporal densities on the extent of disease spread, pesticide use and farmers income,
we compared the effects of the 10 scenarios on three criteria:  proportion of infected fields ($I$), proportion of fields treated ($T$) and expected gross margin ($R$). 
Means and proportions were taken over the last four simulated years (after the emergence stage) and over fields in the inner 10 x 10 grid of fields.

\subsubsection*{Sensitivity analysis}
Our simulation study revealed a higher impact of spatial density than temporal density on the number of treatments applied. Since a spatial density of 50\% of the fields is unrealistic in practice (due to costs), we considered PMN3 with $h=1$
as a good PMN candidate and we further studied the  model behaviour in this case.
For each pest type, we studied  which parameters among the pest dynamic parameters and the economic parameters, had the larger influence of  $I$, $T$ and $R$, using Sensitivity Analysis  (SA, see SI for a complete description of the procedure). We focused on the analysis of 6 parameters (called factors of the SA) linked to the type of crop pest (see Table \ref{tab:factors}).
Since simulations are costly over time, we  first built a statistical metamodel with a negligible computational time to approximate the relationship between the  6 factors and the  criteria (sometimes referred to as an emulator, ~\cite{Prowse2016}). We then we used the metamodel to estimate the part of the model variance explained by the pest dynamic parameters and the economic parameters (using Sobol indices \cite{SOBOL2001}) for each pest type. The domains of variation of the factors used to build the metamodel and to compute the Sobol indices are reported in Table \ref{tab:factors}.

\begin{table}[!ht]
\begin{tabular}{lcccc}
\hline
Factor  & Soil-borne pathogens & Weeds  & Insects & Metamodel\\
\hline
$\epsilon$  &  [0.01 0.10] & [0.1 0.2] & [0.25 0.35] & [0.01   0.4]\\
\hline
$\rho$ &  [0.05 0.15] & [0.15 0.25] & [0.35 0.45] & [0.05   0.5]\\
\hline
$\nu$  & [0.45 0.55] & [0.45 0.55] & [0.2 0.3] & [0.10   0.6]\\
\hline
$c_{pest}$  &  [20 60] & [130 170] & [20 60] & [20   200]\\
\hline
$\gamma$ &  [0.7 0.9] & [0.8 0.95] & [0.6 0.8] & [0.6   0.95]\\
\hline
$q$  &  [0.6 0.8] & [0.6 0.8] & [0.7 0.9] & [0.5   0.9]\\
\hline
\end{tabular}
\caption{{\bf Domains of variation of the model parameters considered as factors for the sensitivity analysis.} 
The last column indicates the domains used to build the metamodel, whereas columns 2 to 4 indicate the domains used to compute the Sobol indices of each submodel associated with each pest.}
\label{tab:factors}
\end{table}

The code of the complete study (PMN comparison and SA) is available on figshare (doi: 10.6084/m9.figshare.7583258.v1).

\section{Results}

\begin{figure}[!h]
\centering Soil-borne pathogens\\
        \includegraphics[width=4.3cm]{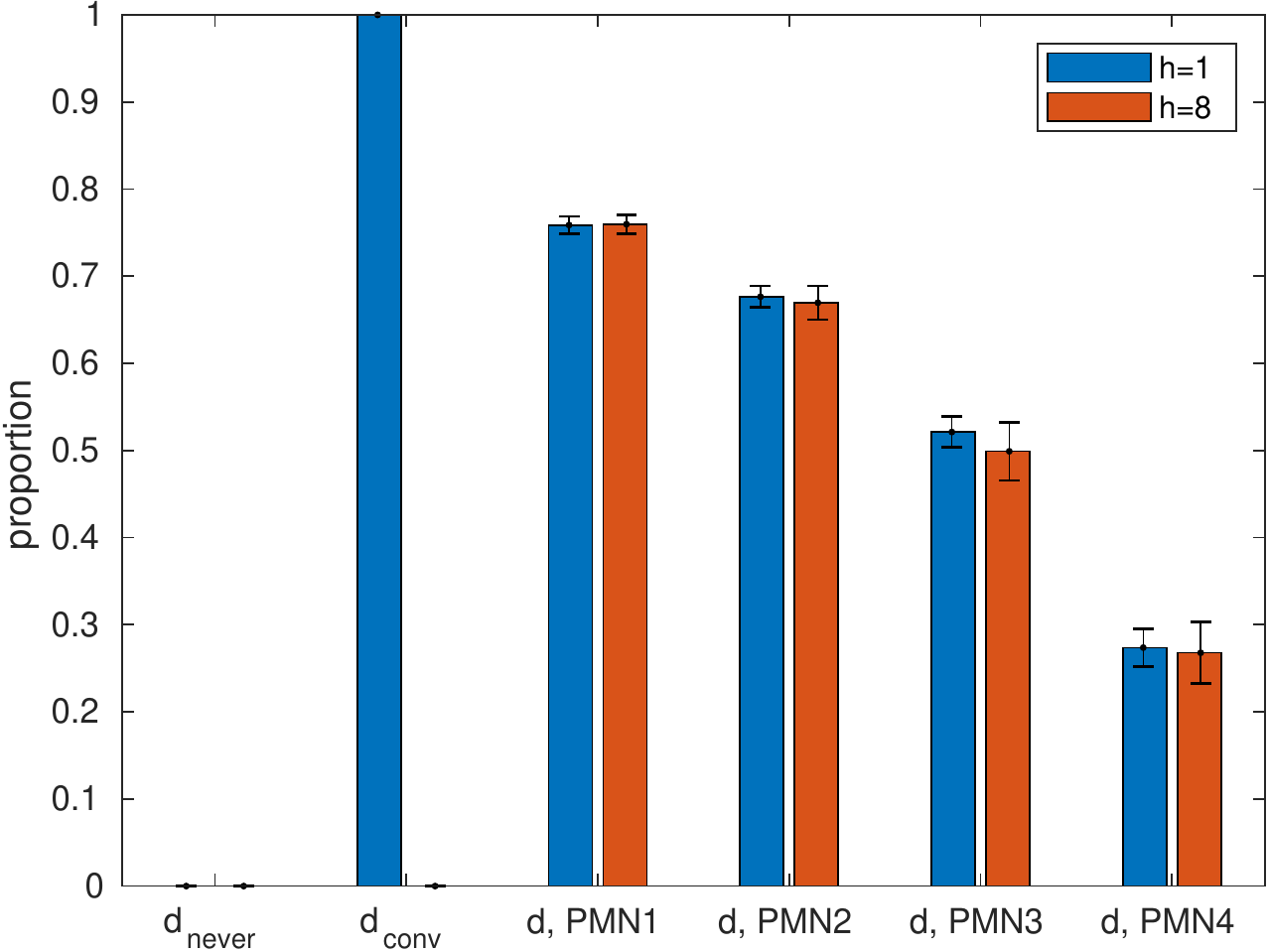}
        \includegraphics[width=4.3cm]{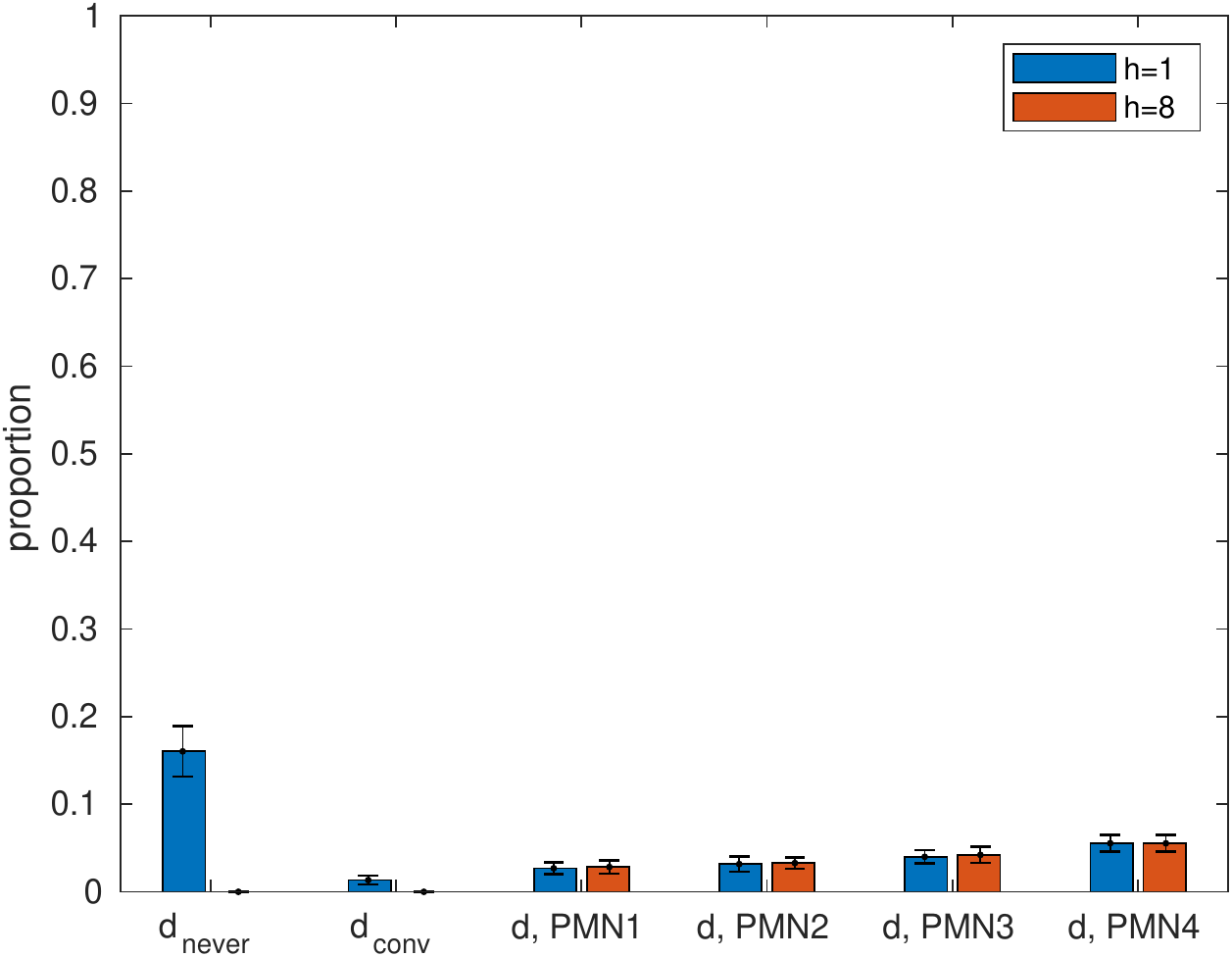}
        \includegraphics[width=4.3cm]{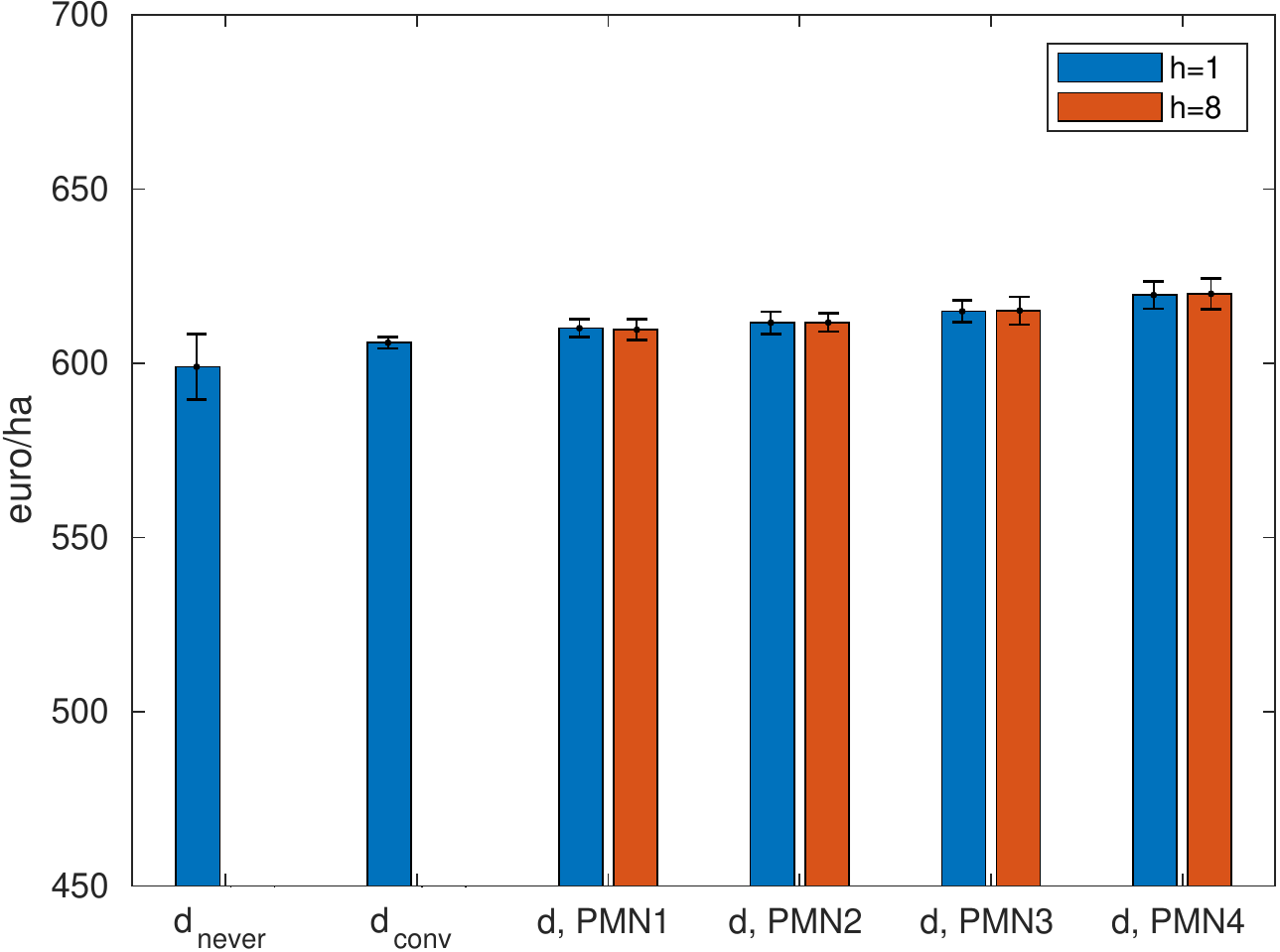}\\
\centering  Weeds \\
        \includegraphics[width=4.3cm]{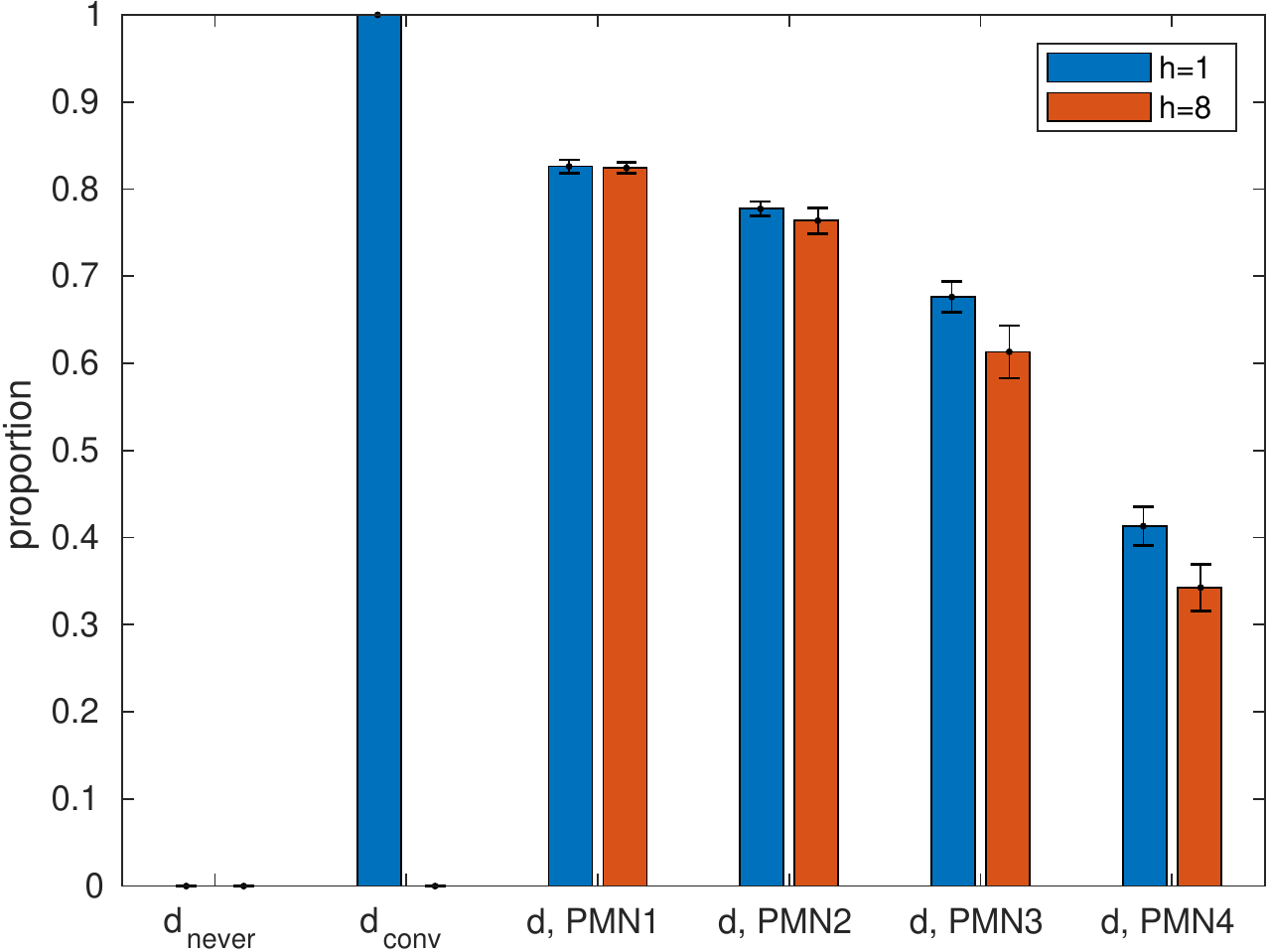}
        \includegraphics[width=4.3cm]{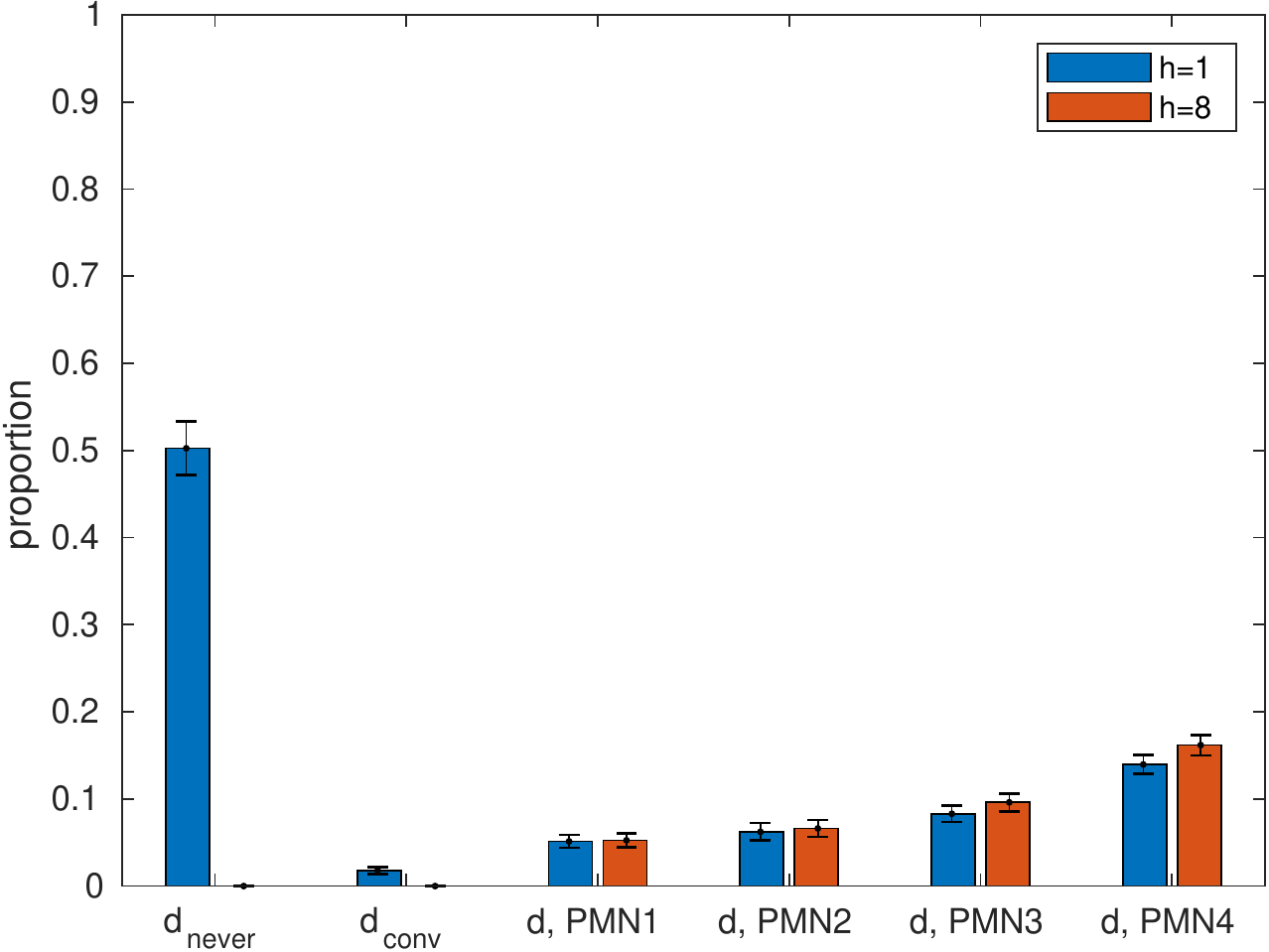}
        \includegraphics[width=4.3cm]{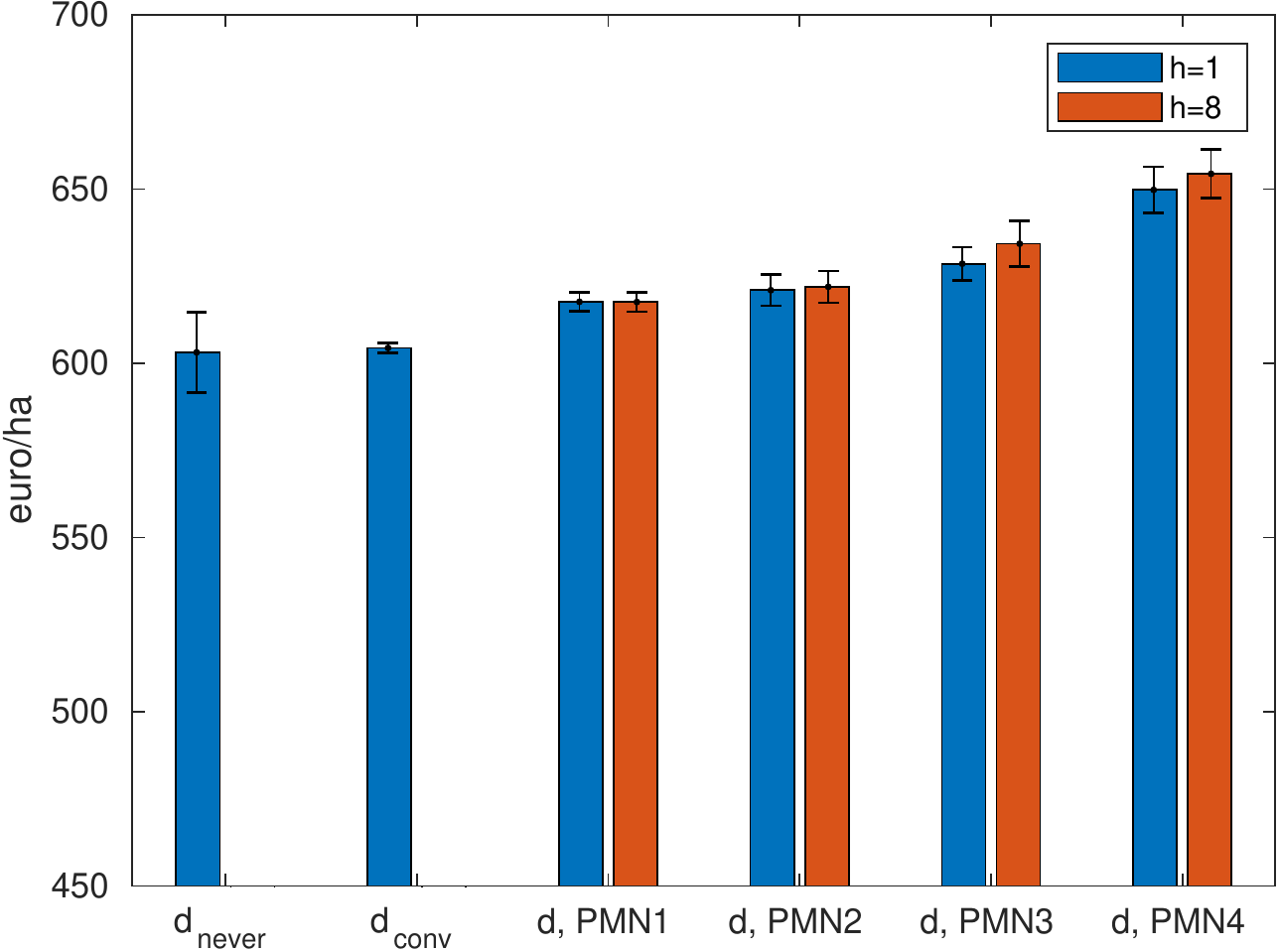}\\
\centering  Insects \\
        \includegraphics[width=4.3cm]{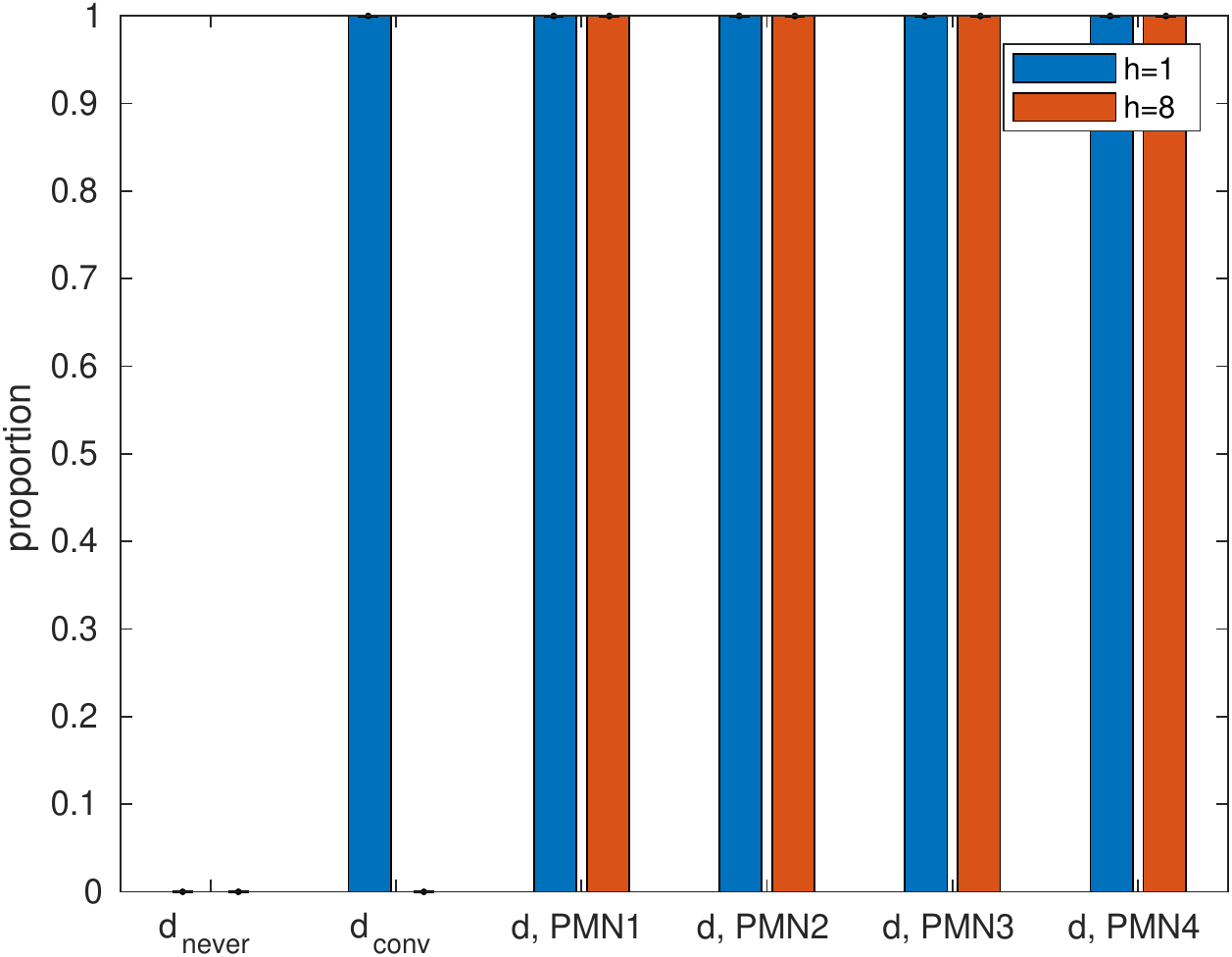}
        \includegraphics[width=4.3cm]{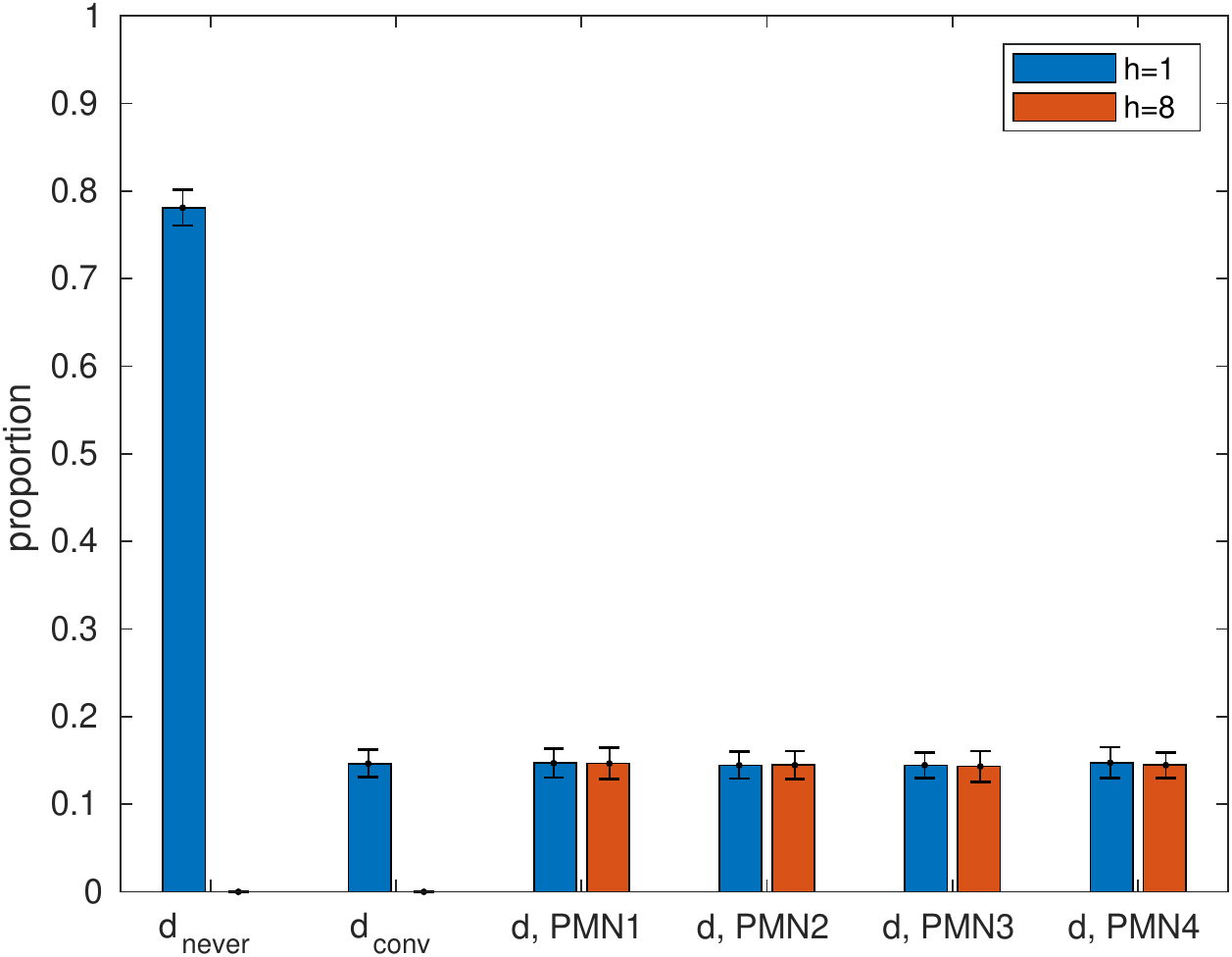}
        \includegraphics[width=4.3cm]{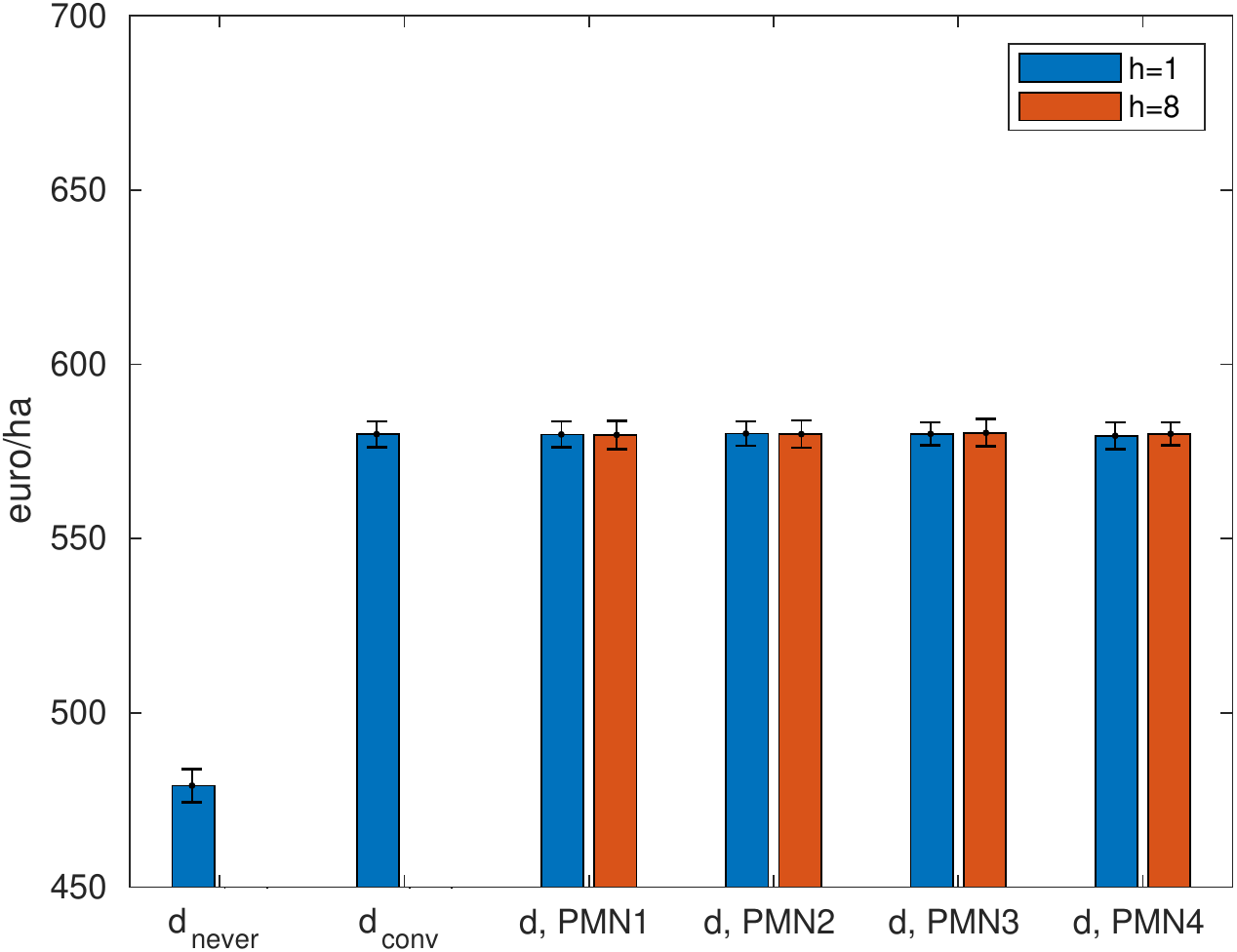}\\
        \centering{T \hspace{3.9cm}I \hspace{4.1cm}R}\\
  \caption{\textbf{Influence of PMN spatial and temporal densities on pest management.}\\
In each graph, the scenarios are represented on the horizontal axis: from left to right, never treat (rule $d_{never}$), conventional treatment (rule $d_{conv}$), then rule $d$ with PMN1 to PMN4 and for those cases with two history depths, $h=1$, $h=8$. 
  Left column: proportion of treatment decision ($T$); middle column: proportion of infected fields ($I$); right column: mean gross margin ($R$).
Vertical error bars represent the standard deviation.}
\label{fig:ITR_K1}
\end{figure}

We observed the same behaviour of the three criteria on simulations for soil-borne diseases and weeds (see figure~\ref{fig:ITR_K1}).

The proportion of treatment applications $T$ decreased when spatial and temporal network densities increased
but with a much stronger effect of the network spatial density. Considering the rule $d$ (the one using PMN and private information) with temporal density $h=1$, $T$ decreased from 75.9\% for PMN1 (1\% of fields observed) to 27.4\% for PMN4 (50\% of fields observed) for soil-borne diseases, and from 82.6\% to 41.3\% for weeds. Considering rule $d$ and PMN4, $T$ decreased from 27.4\% for $h=1$, to 26.8\% for $h=8$ for soil-borne diseases and from 41.3\% to 34.3\% for weeds.  

The mean gross margin increased when spatial and temporal network densities increased, in particular for PMNs 
with larger spatial density (especially for weeds because their annual treatment cost is much higher). 
Considering rule $d$ and $h=1$, 
$R$ increased from {\euro}610/ha for PMN1 to {\euro}620/ha
(+{\euro}10/ha) for PMN4 for soil-borne diseases, and from {\euro}618/ha to {\euro}650/ha (+{\euro}32/ha) for weeds. 
Considering rule $d$ and PMN4, 
$R$ increased from {\euro}619/ha for $h=1$ to {\euro}620/ha (+{\euro}1/ha) for $h=8$ for soil-borne diseases and from 650 to {\euro}654/ha (+{\euro}4/ha) for weeds.

Although injury severity slightly increased with larger PMNs, due to a decrease in pesticide use, 
soil-borne diseases and weeds remained under control. Considering rule $d$ and $h=1$, $I$ increased from 2.7\% for PMN1 to 5.5\% for PMN4 for soil-borne diseases and from 5.1\% to 14.4\% for weeds. Considering rule $d$ and PMN4, $I$ is nearly constant for $h=1$ and $h=8$ for soil-borne diseases and increased from 14.0\% for $h=1$ to 16.2\% for $h=8$ for weeds.

Moving from extreme rule $d_{never}$ (no treatment) to rule $d$ with $h=8$ and PMN4, 
the average gross margin slightly increased and the injury severity decreased.
$T$ reached 26.8\% for soil-borne diseases and 34.3\% for weeds. $I$ decreased from 16.0\% to 5.5\% for soil-borne diseases, and  from 50.3\% to 16.2\% for weeds. $R$ increased from {\euro}599/ha to {\euro}620/ha (+{\euro}21/ha) for soil-borne diseases and from {\euro}603/ha to {\euro}654/ha (+{\euro}51/ha) for weeds. 

Moving from conventional rule $d_{conv}$ (always apply conventional treatment) to rule $d$, $h=8$, PMN4, 
the average gross margin slightly increased and the injury severity increased.
$T$ decreased to 26.8\% for soil-borne diseases and to 34.3\% for weeds. 
$I$ increased from 1.3\% to 5.5\% for soil-borne diseases and remained roughly constant for weeds. 
$R$ increased from {\euro}606/ha\ to {\euro}620/ha (+{\euro}14/ha) for soil-borne diseases and from {\euro}604/ha to {\euro}654/ha (+{\euro}50/ha) for weeds.

The results for insects were quite different (see figure~\ref{fig:ITR_K1}). The strategy, which consists in never using treatments, led to 78.1\% of infected fields ($I$), a value much higher than for soil-born pathogens (16.0\%) and weeds (50.3\%). This is due to the high dispersal of insects at short and long distances.
With rule $d$, treatment decision was always chosen regardless of the PMN considered because the probability of infection of a field estimated from the PMN observations, $p_i^t$, was always higher than the decision rule threshold. Information from the PMN  was not sufficient to decrease treatment frequencies because treatment cost is low, which implies that the threshold is low.

For rule $d_{never}$ (no treatment), the mean gross margin is reduced ({\euro}479/ha) while it is quite constant otherwise ({\euro}580/ha, see figure~\ref{fig:ITR_K1}).

These results were obtained for a decision rule $d$ where only the private information of the previous year was used.
When we increased the history depth of the private information using the three previous years 
 to determine the treatment threshold in the decision rule $d$, we  observed the same tendencies  but mitigated (see figure~\ref{fig:ITR_K3} in Appendix B). However, the expected gross margin was larger in the case of soil-borne pathogens and pests, and treatment was no longer systematically applied for insects ($T$ = 88\%).

Focusing now on PMN4 with $h=1$ as a good compromise between monitoring efforts and  treatments impact,  the sensitivity analysis made it possible to evaluate the influence of some parameters on the model behaviour. The analysis performed showed that regardless of the pest type, varying the long-distance dispersal ($\epsilon$) and the probability of persistence ($\nu$) had little influence on $I$, $R$ and $T$ (Sobol index estimators are reported in Table \ref{tab:sobol}). 

Consequently, for soil-borne pathogens and weeds, the most influential parameters on $I$ and $T$ are the annual cost of specific treatment ($c_{pest}$), and the yield when infected ($q$), whereas the main influential parameters on $R$ are the probability of  contamination from neighbouring fields ($\rho$) and  $q$.

For insect pests, the factors that have the most influence on $I$ are $q$ and the probability of treatment efficacy ($\gamma$); for $T$, they are $q$ and $c_{pest}$; and for $R$, it is $\rho$.

\begin{table}[!ht]
\begin{tabular}{lcccccc}
\hline
\textbf{Prop. of infected fields, $I$}   & $\epsilon$  & $\rho$ & $\nu$ & $c_{pest}$ & $\gamma$ & $q$ \\
\hline
Soil-borne pathogens &  0.06 &  \colorbox{yellow}{0.27} &  0.03 &  \colorbox{orange}{0.54} &  0.02 &  \colorbox{yellow}{0.32} \\
\hline
Weeds & 0.05 &  \colorbox{yellow}{0.28} &  0.03 &  \colorbox{orange}{0.53} &  0.02 &  \colorbox{yellow}{0.31} \\
\hline
Insects &  0.10 &  \colorbox{yellow}{0.21} &  0.02 &  0.16 &  \colorbox{orange}{0.48} &  \colorbox{red}{0.69} \\
\hline

\hline
\textbf{Prop. on treatment, $T$}  & $\epsilon$  & $\rho$ & $\nu$ & $c_{pest}$ & $\gamma$ & $q$ \\
\hline
Soil-borne pathogens &  \colorbox{white}{0.04} &  0.03 &  0.01 &  \colorbox{orange}{0.41} &  0.15 &  \colorbox{orange}{0.41} \\
\hline
Weeds &  0.03  & 0.03 &  0.01 &  \colorbox{orange}{0.40} &  0.12 &  \colorbox{yellow}{0.37}\\
\hline
Insects &  0.03 &  0.02  & 0.00  & \colorbox{yellow}{0.30} &  0.13 &  \colorbox{orange}{0.48}\\
\hline

\hline
\textbf{Mean net  margin, $R$} & $\epsilon$  & $\rho$ & $\nu$ & $c_{pest}$ & $\gamma$ & $q$ \\
\hline
Soil-borne pathogens &  0.14 &  \colorbox{orange}{0.45} &  0.01 &  \colorbox{yellow}{0.23} &  \colorbox{yellow}{0.29} &  \colorbox{red}{0.60}\\
\hline
Weeds & 0.14 &  \colorbox{orange}{0.45} &  0.01 &  \colorbox{yellow}{0.23} &  \colorbox{yellow}{0.30} &  \colorbox{orange}{0.57} \\
\hline
Insects & 0.07 &  \colorbox{violet}{0.86} &  0.08 &  0.00 &  0.00 &  0.00\\
\hline
\end{tabular}
\caption{{\bf Sobol indices for  proportion of infected fields, proportion of treatment and mean  gross margin in the case of PMN3 and a temporal density $h=1$.} The background colours of indices (white, yellow, orange, red, purple) 
represent the increasing importance of Sobol indices.}
\label{tab:sobol}
\end{table}

\section{Conclusion and Discussion}

In this study, we aimed at investigating the structure of Pest Monitoring Networks (PMNs) that provide
information meaningful enough to control pests and reduce pesticide use. Using a 
stochastic model, we explored eight PMN spatio-temporal structures and examined their 
efficiency to control three pest families that differed by their ability to 
disperse. PMN efficiency was evaluated using three criteria: proportion of infected fields, proportion of fields treated, and the economic benefit to the farmer.

We proposed a Dynamic Bayesian Network model of development and eradication of 
several types of pests at the landscape level.
The model includes the design of an expected margin-based decision rule that 
takes  account of both PMN information and private information about the 
decision-maker's field in order to choose whether to apply treatment actions.

Extensive simulations of the model allowed us to highlight the impact of the 
spatial density of a PMN and the length of considered histories on the 
efficiency and the cost of the control of soil-borne diseases and weeds.
Indeed, the study shows (as expected) that densifying a PMN spatially and 
temporally helps to better control pests with less pesticides. 
We found that for soil-borne pathogens and weeds, increasing the spatial density of a PMN made it possible to significantly decrease the number of treatments (up to 67\%), with only marginal increased infection. 
Considering past observations only had a second-order effect (up to 13\%). This could seem contradictory with some recent studies \cite{Holden2016} where  a long-term surveillance effort was recommended (possibly with a varying effort). However, our results do not indicate that surveillance should only be restricted to short periods but, instead that each year, a good decision on whether to treat or not only require that the decision-maker be informed by the past short-term surveillance data.

The PMN had almost no impact on treatment reduction in the case of  insects, except when private information was considered (20\% decrease in treatments). This may be due to the spatial scale of our simulation study, whose extent is too small compared to the dispersion capacity of insects. 
For the two other pests,  dispersion is  low and local information at the field scale in enough to control the epidemics. However, insects disperse too fast and are likely to very rapidly infect the whole area considered if not treated. Thus, in our  model, it is likely that for insects, treatment decisions should be decided at a larger scale (several thousand  hectares) on the basis of information about fields pcarsely distributed in the region, rather then decided at as small a scale like for the two other pathogens.

Even though the interest of a PMN has been highlighted for weeds and soil-borne pathogens, the results also show the importance of considering the past private information of a field for treatment decision. In practice this private information is more widely shared than in our model: farmers own more than one field and they exchange information about their fields with other farmers. Therefore the impact of private information is probably even larger than predicted by our model.

Finally, for a given PMN, Sensitivity Analysis (SA) was a way to assess the relative importance of epidemic and economic parameters on the simulation model outputs. 
It allowed  us to identify some important parameters of the model that would influence the treatment intensity induced by a given decision rule.
We identified leverage effects via the factors that have a strong influence on the proportion of infected fields, the proportion of treatments and the mean net margin. They are different for insects and for weeds or soil-borne pathogens. For example, modifying the cost of pesticides may have an impact on the number of treatments and of infected fields for the latter. On the contrary, for insects, these two outputs are more highly influenced by the remaining yield after infection, which is not a factor that humans can easily modify.
SA was also a way to evaluate the model's robustness: the proportion of infected fields, the proportion of fields treated and the mean gross margin have a low sensitivity to the variations of some of the model parameters. This is true, in particular, for the strength of long distance dispersal and the probability of survival in the case of the three pest types. This is consistent with the fact that primary inoculum is generally considered as non-limiting in epidemiological models \cite{Madden2007}.
These results are obtained for a PMN covering 25\% of the fields and with a history of length 1 (only  observations from the previous year). They may differ for another PMN structure and for another decision rule.

A next step for concretely designing efficient PMNs would be to look for a PMN that minimises three objectives: pesticide use, crop losses caused by pests, and costs. 
Since we have a stochastic simulation model for assessing PMNs, it would be natural to apply simulation optimization approaches
\cite{FU15}. However this remains a complex multi-criteria optimization problem.

Beyond PMN analysis or design, our model can also serve as an interactive tool for discussion between farmers and decision-makers to better understand the spatial and temporal connections underlying pest dynamics and to help design management strategies at the landscape level~\cite{Debaeke2019} and to identify the most promising ones. 
To do so, the model should be extended in two directions to increase its representativity.
 The first one would be to include more heterogeneity in the model's components. The second one would be to enrich management options.
Regarding the first direction,  only a single crop is currently considered throughout the landscape for the simplicity 
of interpretation. This may be true only in some specific agricultural areas. We shall first include landscape heterogeneity by 
incorporating several crops (wheat, barley, etc) with different sensitivities to and hosting capacities for pests, as well as several crop rotations. In order to  adapt to each region's specificity,
the simulator could also easily be extended to take  information about   climate variations (successions of dry and humid years, warm and cold years) into account and/or  the evolution of  economic parameters evolution over time.
In the same way, 
our model could include spatial heterogeneity, not only in terms of crop diversification, but also in terms of interstitial spaces (e.g. field edges, hedgerows) and semi-natural habitats (e.g., deciduous woodlands). It is well-known that diversified crop sequences (and intercrops) disrupt the cycles of soil-borne diseases. Similarly, alternating sowing periods makes it possible to limit weed specialisation in a given field. As for animal pests, it is considered that the planned biodiversity will increase the associated biodiversity, including natural enemies, that can limit pest development. A substantial improvement of the approach presented in this paper could be to take  landscape heterogeneity into account as an important driver of pest development.

In the second direction, we shall enrich the PMN-based pest-management decision rule.
A decision rule is composed of a set of available actions among which the decision-maker must choose, and a criterion (the expected margin in this case) that will be maximized by the action choice.
We could consider another criterion than the expected margin
 to take  account of the fact that the decision-maker could possibly be risk-averse.
Indeed, it has been  experimentally verified for a long time now that even though risk-aversion has an impact on pesticide use by farmers \cite{Pannell1991}, this impact depends on the  parameters considered and especially on expected yields, and may not always result in an increase in pesticide use \cite{Aka2018}. 
Our model should therefore be of use to test the plausibility of different risk-averse treatment decision rules.
 
Oreover, the current computation of the expected gross margin relies on the assumption that the costs to build and manage the PMN are paid by a third party such as a human health care entity or a  drinking water company. However, if the PMN information can benefit  the farmers, it could  be envisaged that the PMN costs could be shared between them.
It would be interesting to see how the results of the study are modified in this case. In the same way, the decision to treat could be deeply affected by accounting for the hidden costs attached to pesticides, notably, decontamination costs \cite{Bourguet2016}. In addition,  pesticide reduction would contribute to the limitation of biodiversity loss and human health hazards. 

More complex actions, beyond treatment/no treatment, could also be
considered, by combining several types of actions. This would better represent the situation that farmers or decision-makers are faced with.
PMN information and treatment choice could be combined with other actions including agricultural practices that reduce biotic risks or that  increase treatment efficacy.
For example, in \cite{Cros2017}, pest-control through crop rotation and cultivar choices was optimized at the landscape scale. A stochastic model of pest dynamics was used, that was very similar to the one we used in this study. For example, \cite{Pelzer2010} developed a spatially explicit model to manage phoma stem canker at the landscape level. This model represents the effects of many cropping practices in interaction with weather scenarios: crop sequence, tillage, nitrogen management at the cropping system level, cultivar choice, sowing date and rate, and fungicide treatments.
The choice of a combined action could then  be taken yearly or for a sequence or years (multi-year pest-management strategies, such as crop rotations). It could be made independently in each field, 
or as the result of a coordinated  decision between decision-makers. 
Finally, the decision rule could target several pests at the same time since some treatments are  multi-purpose.

As long as  the PMN influence is explored through simulation, all these enhancements are easy to integrate into the model. 
The model that we proposed here can be seen as the first step in building richer simulators of pest dynamics and their control based on PMN observations.

\vspace{0.3cm}

\section*{Acknowledgements}
We thank Victor Picheny and Robert Faivre of MIAT INRA Toulouse, France for their help in the choice of the sensitivity analysis procedure and the interpretation of results.
This work was partly supported by the ECOPHYTO project VESPA.

\bibliography{biblioVESPA.bib}
\bibliographystyle{vancouver}

\section*{Appendix A. Parameter values}

\begin{table}[!ht]
\begin{tabular}{l l c c l}
 \hline
 \rowcolor{lightgray}
 Parameter  & Definition &  Unit & Value & Reference \\
 \hline
$yield$    & maximal annual yield                               & kg/ha   & 3500 & ~\cite{FOP2015,Cros2017}     \\
$price$    & selling price                       & {\euro}/kg& 0.35 & ~\cite{FOP2015,Cros2017}     \\
$c$        & all production costs  (seed, fertilizer,  & {\euro}/ha&  849.4 & ~\cite{FOP2015,Inosys2015,Cros2017}     \\
           & work carried out by third parties, &            &      & \\
           & phytosanitary treatments) &            &      & \\
$c_{pest}$    & cost for fungicide products         & {\euro}/ha& 38   &   ~\cite{Inosys2015} \\
    & cost for herbicide products         & {\euro}/ha& 147  &   ~\cite{Inosys2015} \\
     & cost for insecticide products       & {\euro}/ha& 32   &   ~\cite{Inosys2015} \\
$\epsilon$ & long-distance dispersal probability & &  & \\
 &  for soil-borne pathogens & - & 0.05 & expertise \\
 &for weeds  & - & 0.15 & expertise \\
&for insects  & - & 0.30 & expertise \\
$\rho$   & spreading probability from neighbouring fields & &  & \\
   & for soil-borne pathogens & - & 0.10 & expertise \\
   &  for weeds  & - & 0.20 & expertise \\
  & for insects & - & 0.40 & expertise \\
$\nu$  & probability of pest survival if not treated     & &  & \\
   & for soil-borne pathogens  & - & 0.50 & expertise \\
    &  for weeds  & - & 0.50 & expertise \\
    &  for insects  & - & 0.25 & expertise \\
$\gamma$ & probability of treatment efficacy     & &  & \\
 & for soil-borne pathogens & - & 0.8 & expertise \\
& for weeds  & - & 0.9 & expertise \\
 & for insects & - & 0.7 & expertise \\
$q$ & when infection, proportion remaining of yield    & &  & \\
     & for soil-borne pathogens  & - & 0.7 & expertise \\
     & for weeds & - & 0.7 & expertise \\
     & for insects & - & 0.8 & expertise \\
 \hline
 \\
\end{tabular}
\caption{{\bf Values for the parameters of the pest dispersal model and the decision rule for the case of oilseed rape in France.}}
\label{tab:PestParameters}
\end{table}

\newpage
\section*{Appendix B. Simulation results when using the last three years for private information}

\begin{figure}[!h]
\centering Soil-borne pathogens\\
        \includegraphics[width=4.3cm]{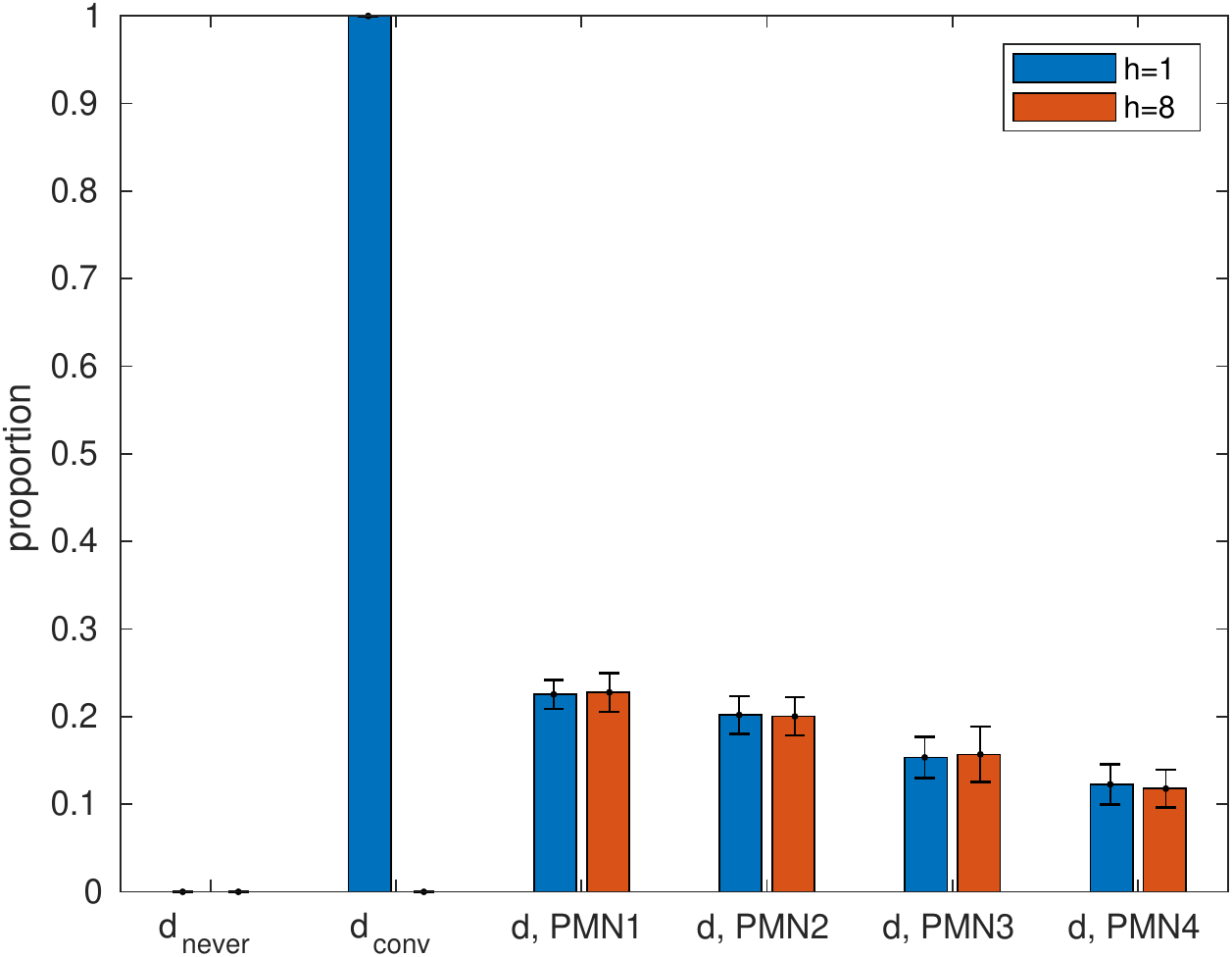}
        \includegraphics[width=4.3cm]{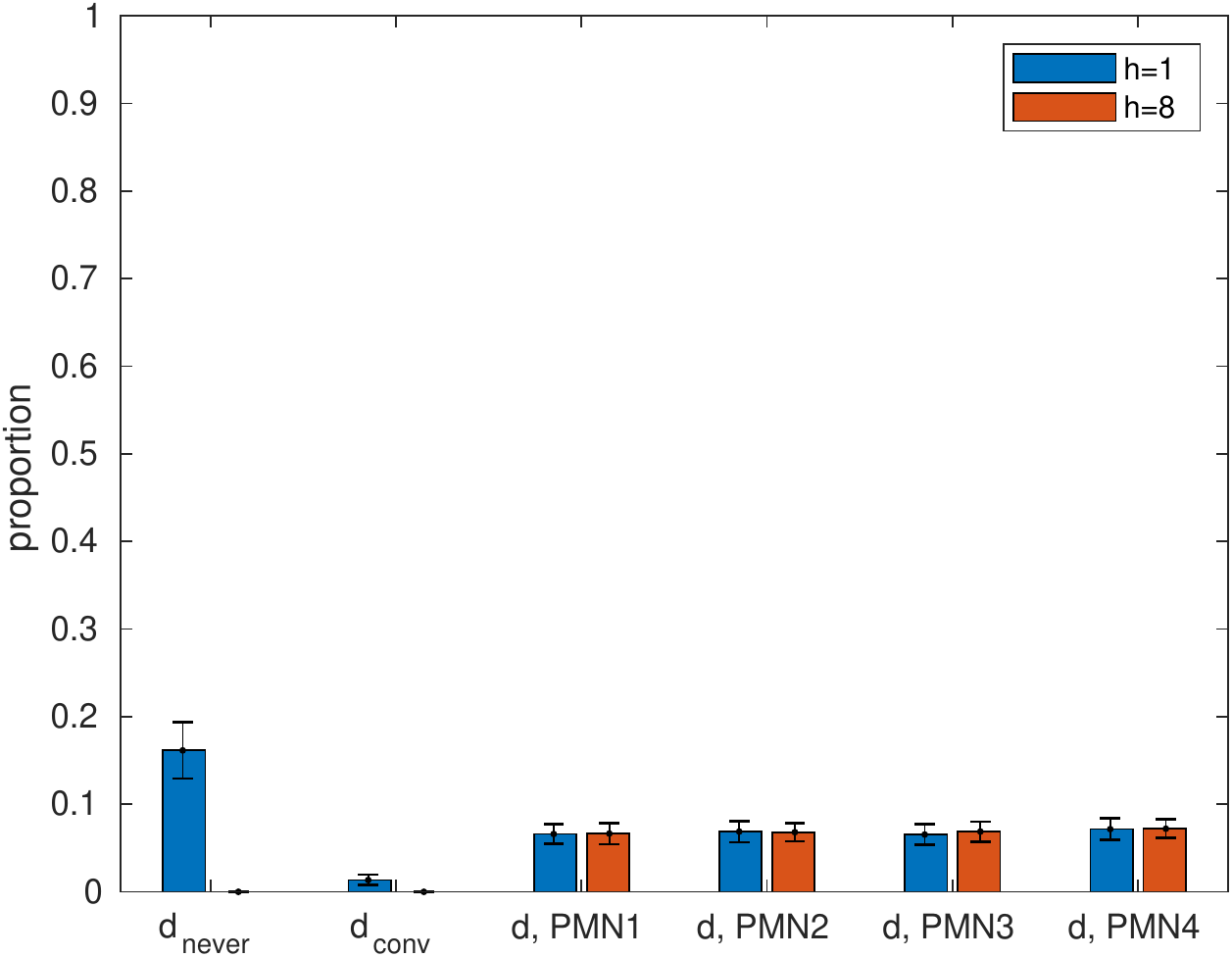}
        \includegraphics[width=4.3cm]{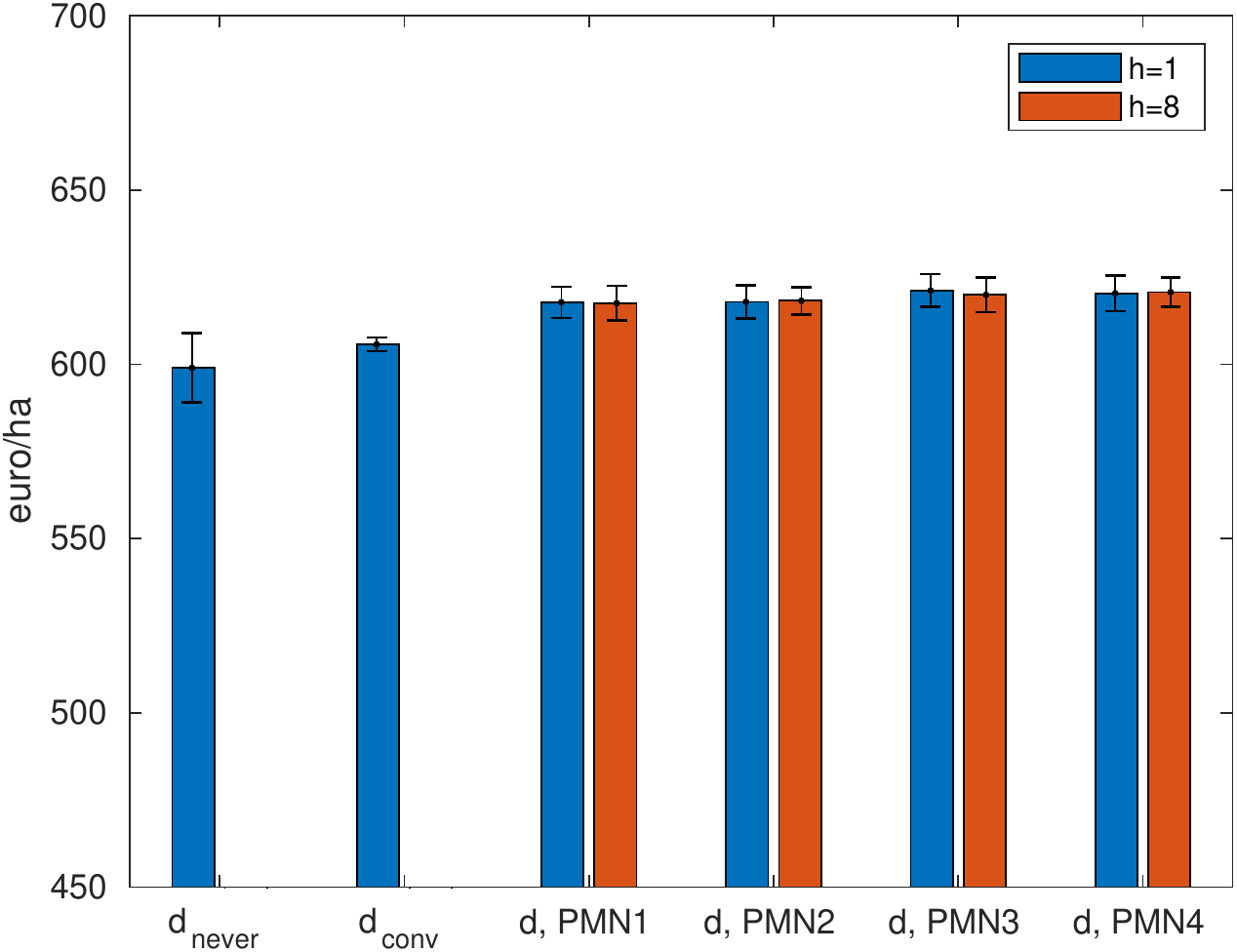}\\
\centering  Weeds \\
        \includegraphics[width=4.3cm]{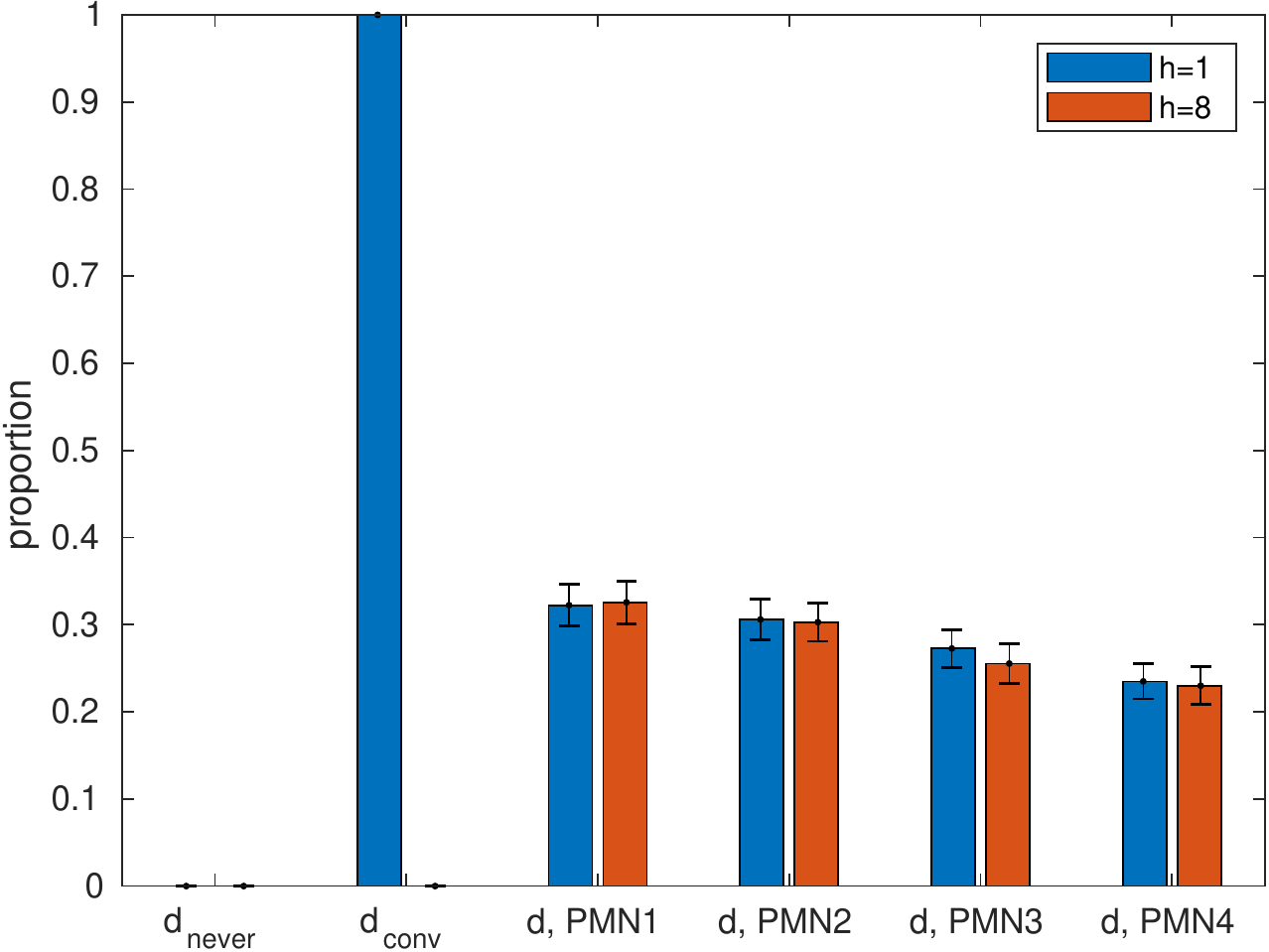}
        \includegraphics[width=4.3cm]{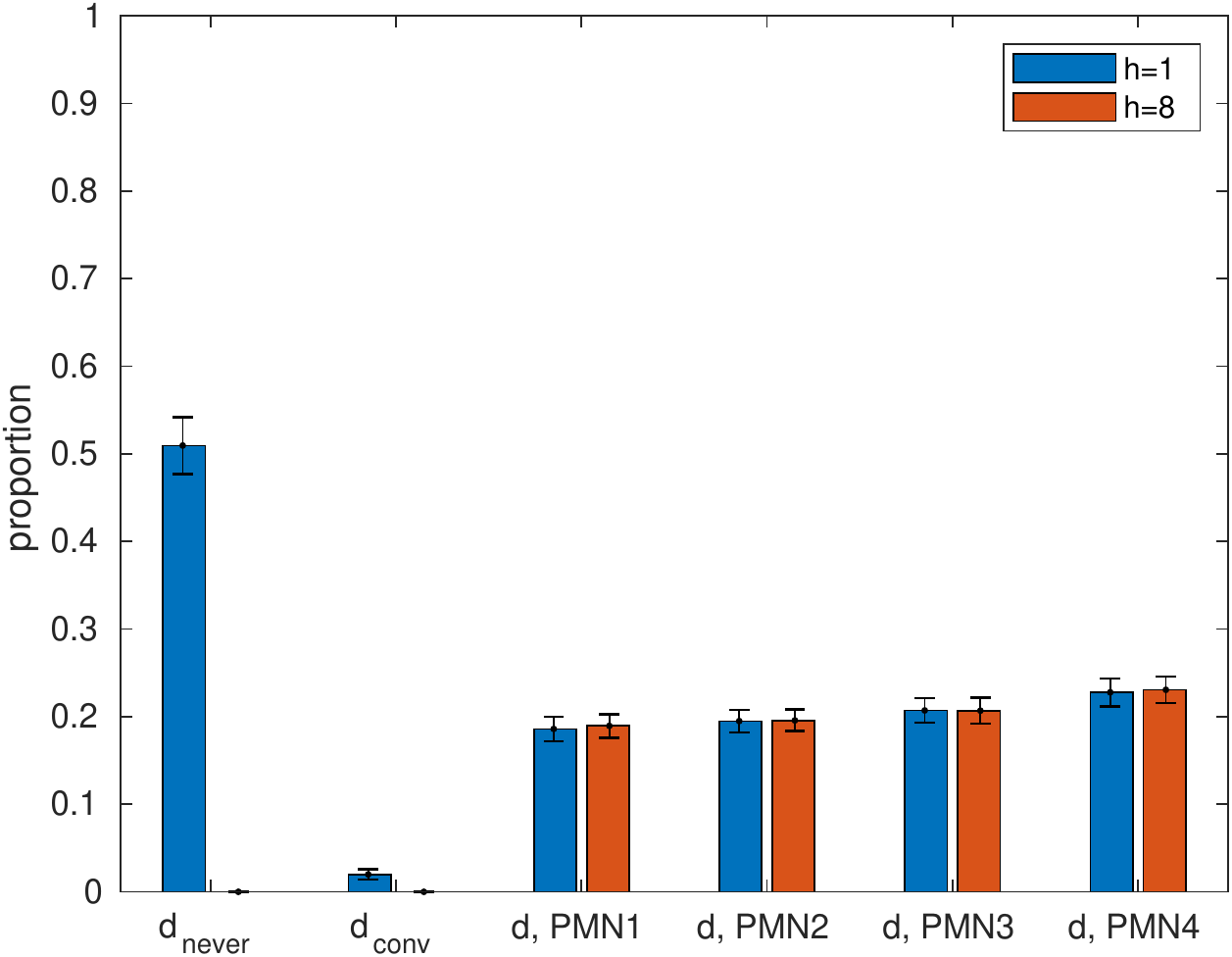}
        \includegraphics[width=4.3cm]{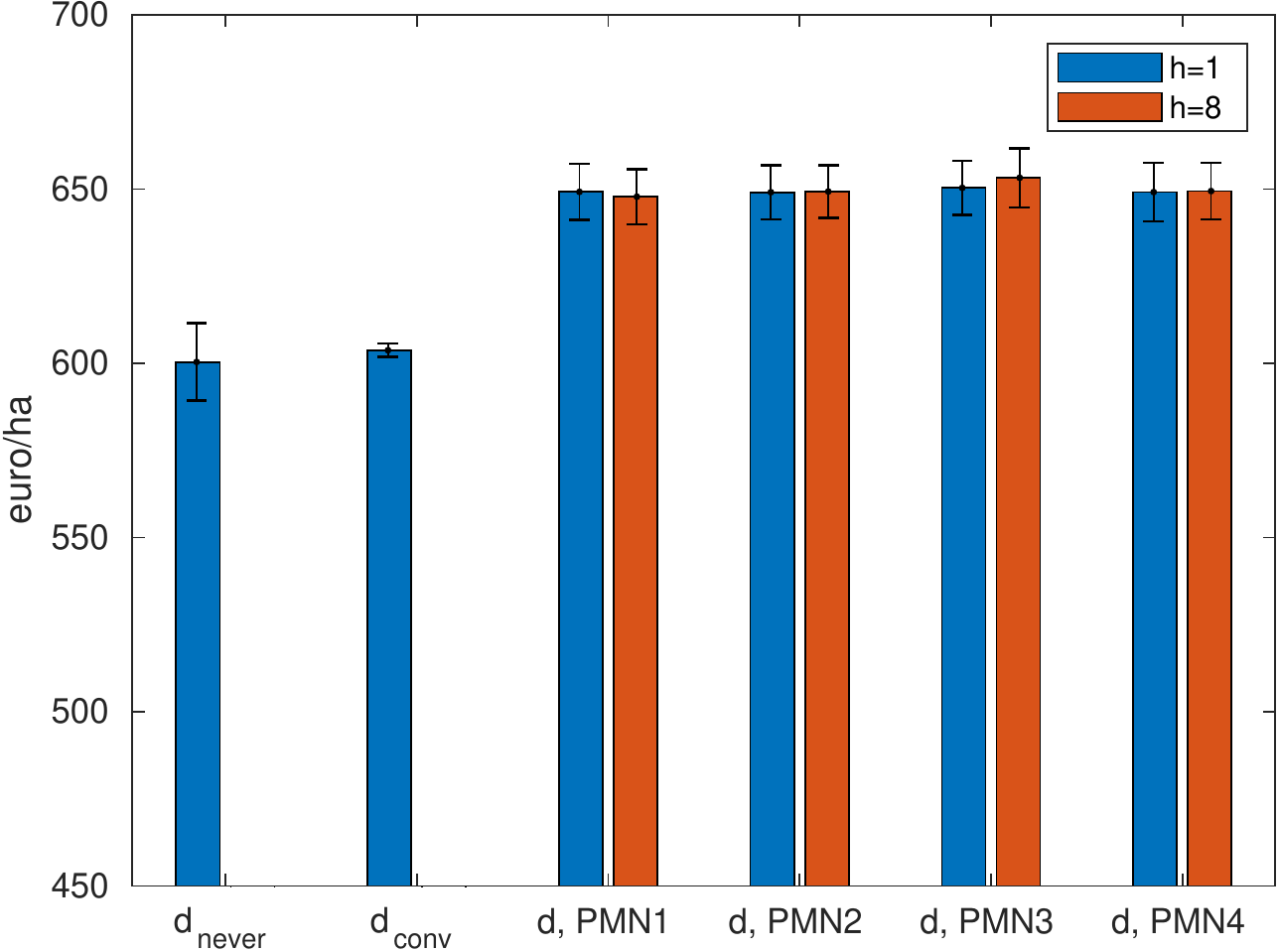}\\
\centering  Insects \\
        \includegraphics[width=4.3cm]{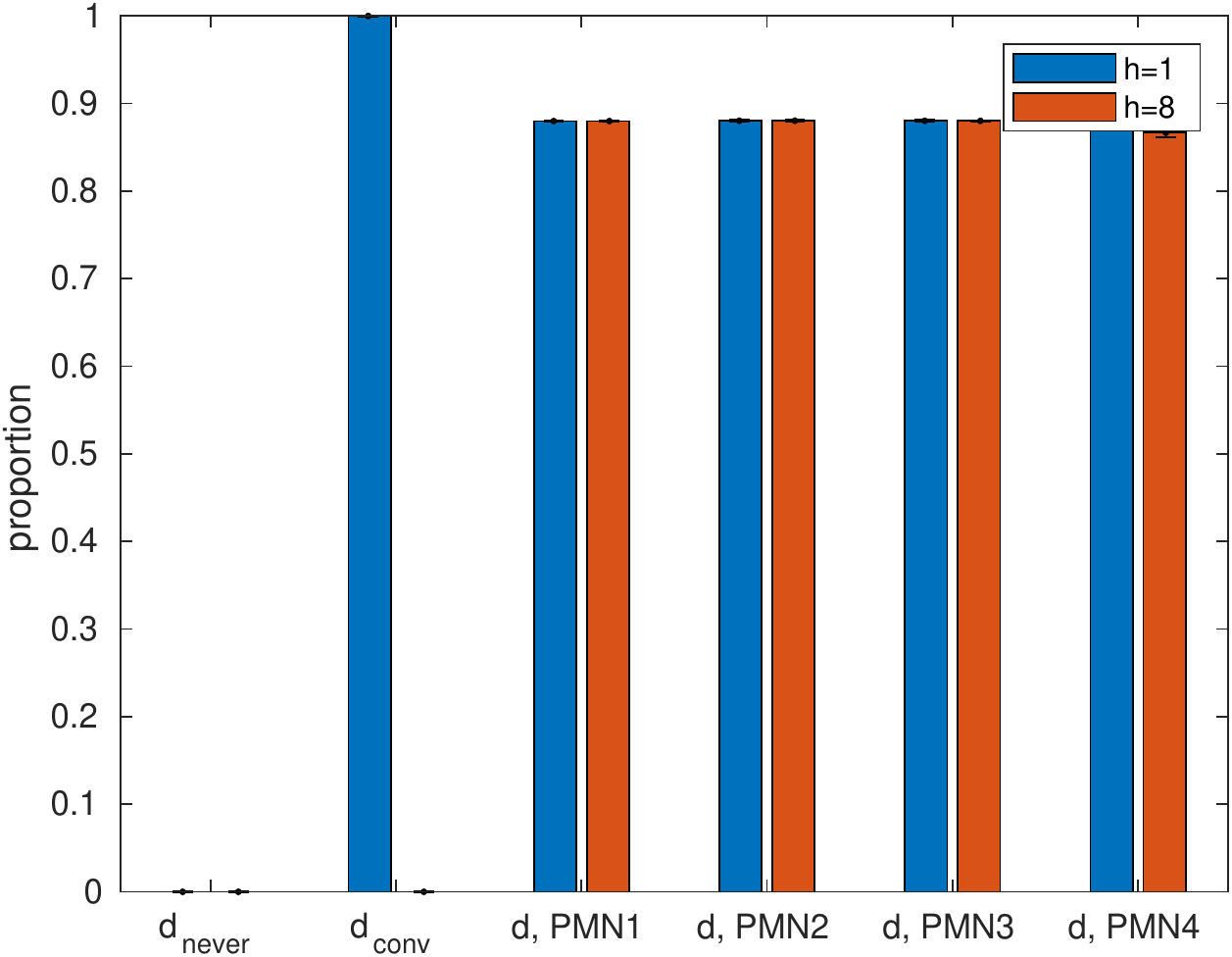}
        \includegraphics[width=4.3cm]{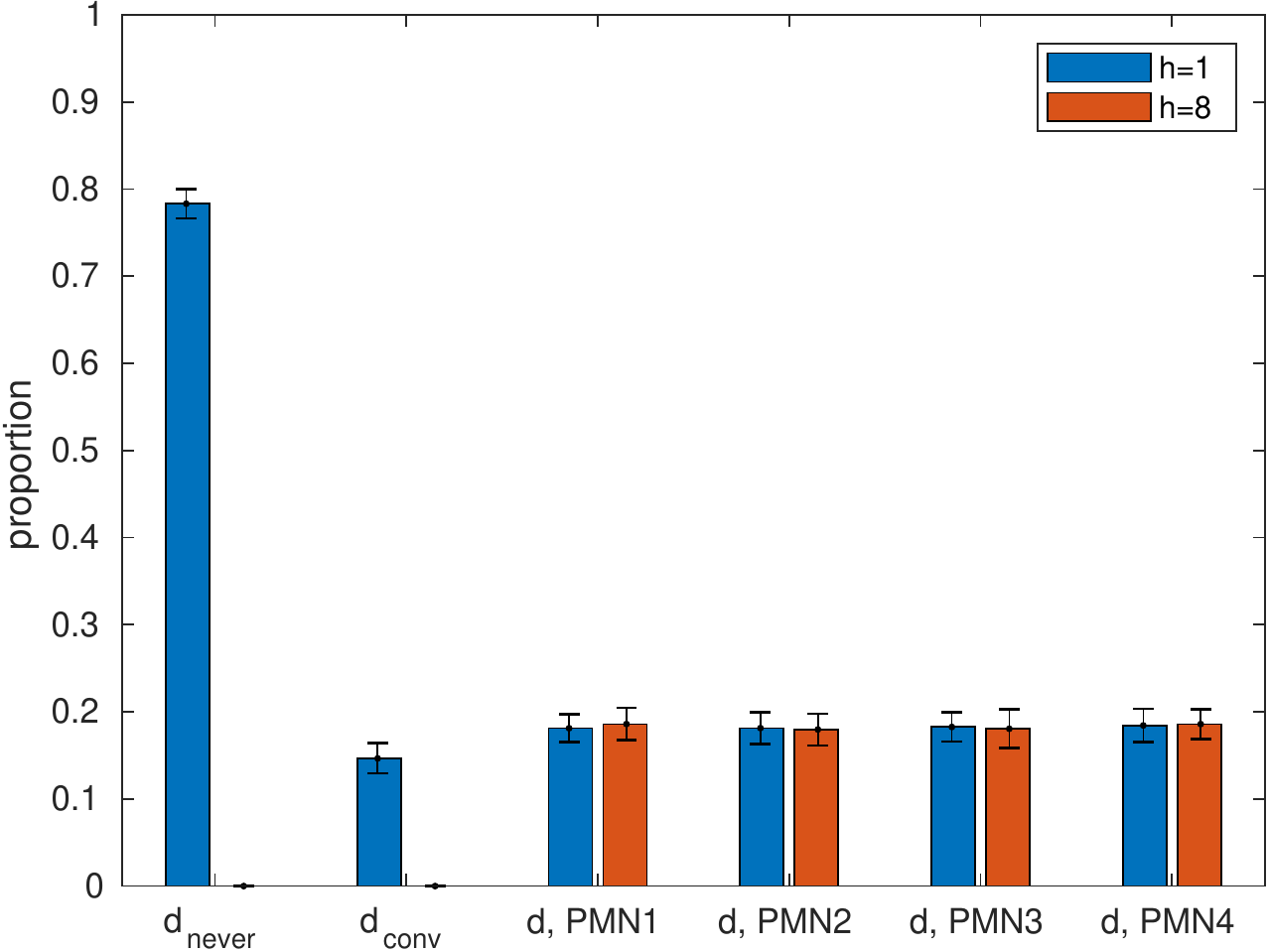}
        \includegraphics[width=4.3cm]{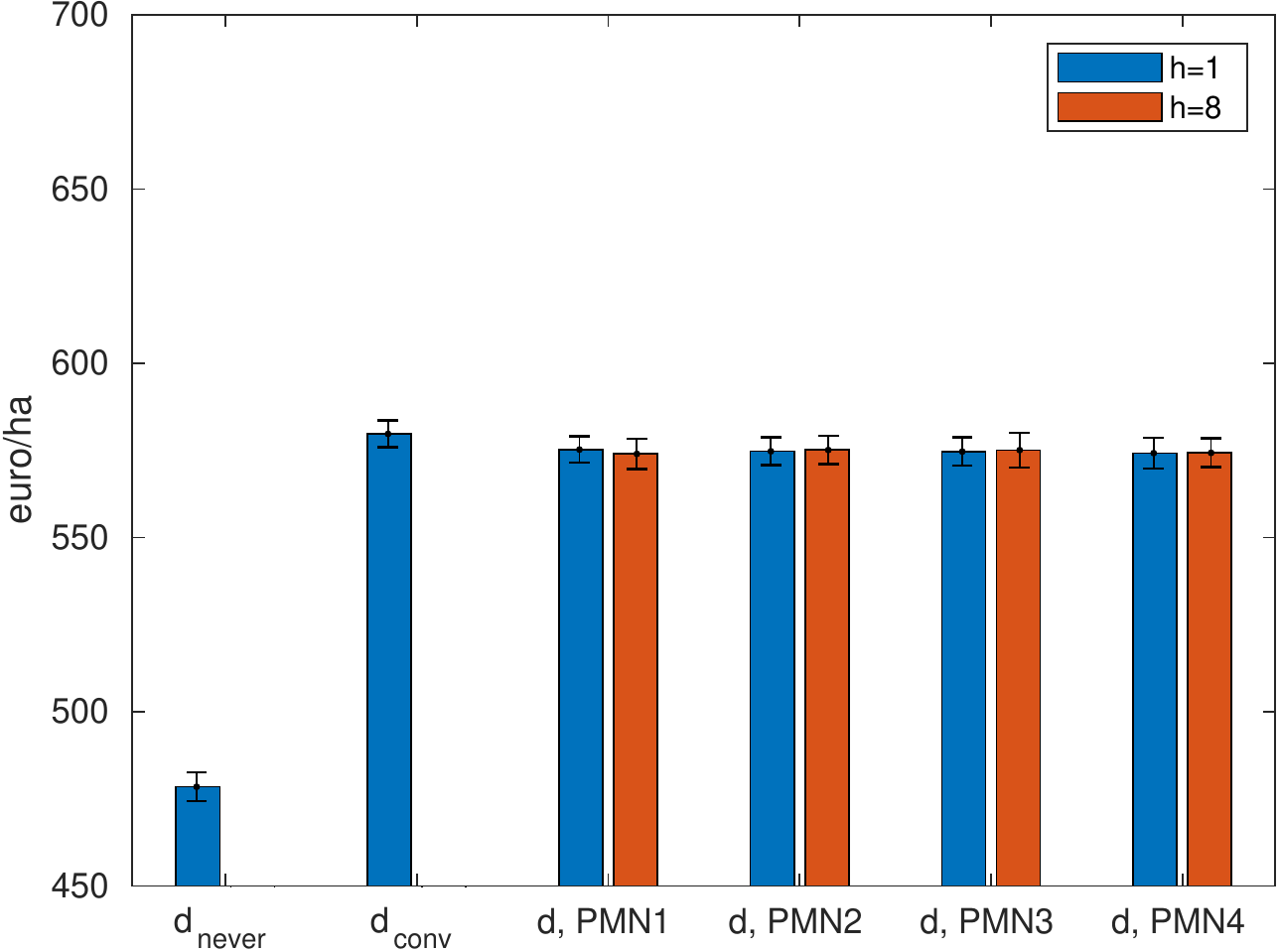}\\
        \centering{T \hspace{3.9cm}I \hspace{4.1cm}R}\\
\caption{\textbf{Influence of PMN spatial and temporal densities on pest management.}\\
In each graph, the scenarios are represented on the horizontal axis. From left to right, never treat (rule $d_{never}$), conventional treatment (rule $d_{conv}$), then rule $d$ with PMN1 to PMN4 and with two history depths $h=1$ and $h=8$. 
  Left column: proportion of treatment decision ($T$); middle column: proportion of infected fields ($I$); right column: mean gross margin ($R$).
Vertical error bars represent standard deviation.}
\label{fig:ITR_K3}
\end{figure}

\newpage

\section*{Supplementary Information}

\subsubsection*{Considered PMN}
\begin{figure}[!h]
\begin{center}
 \begin{tikzpicture}
   \fill[gray!5, pattern=north west lines](0,5.5) rectangle (6,6);  
   \fill[gray!5, pattern=north west lines](0,0) rectangle (6,0.5);  
   \fill[gray!5, pattern=north west lines](0,.5) rectangle (0.5,5.5);  
   \fill[gray!5, pattern=north west lines](5.5,.5) rectangle (6,5.5);  
   \draw [step=.5, very thin, lightgray] (0,0) grid (6,6);
   \fill[gray!60](3,2.5) rectangle (3.5,3);  
 \end{tikzpicture}
\hspace{1cm}
 \begin{tikzpicture}
   \fill[gray!5, pattern=north west lines](0,5.5) rectangle (6,6);  
   \fill[gray!5, pattern=north west lines](0,0) rectangle (6,0.5);  
   \fill[gray!5, pattern=north west lines](0,.5) rectangle (0.5,5.5);  
   \fill[gray!5, pattern=north west lines](5.5,.5) rectangle (6,5.5);  
   \draw [step=.5, very thin, lightgray] (0,0) grid (6,6);
     \fill[gray!60](0+0.5,0.5+0.5) rectangle (0.5+0.5,1+0.5);  
     \fill[gray!60](2.5+0.5,0.5+0.5) rectangle (3+0.5,1+0.5);  
     \fill[gray!60](0+0.5,1.5+0.5) rectangle (0.5+0.5,2+0.5);  
     \fill[gray!60](2.5+0.5,1.5+0.5) rectangle (3+0.5,2+0.5);  
     \fill[gray!60](0+0.5,2.5+0.5) rectangle (0.5+0.5,3+0.5);  
     \fill[gray!60](2.5+0.5,2.5+0.5) rectangle (3+0.5,3+0.5);  
     \fill[gray!60](0+0.5,3.5+0.5) rectangle (0.5+0.5,4+0.5);  
     \fill[gray!60](2.5+0.5,3.5+0.5) rectangle (3+0.5,4+0.5);  
     \fill[gray!60](0+0.5,4.5+0.5) rectangle (0.5+0.5,5+0.5);  
     \fill[gray!60](2.5+0.5,4.5+0.5) rectangle (3+0.5,5+0.5);  
  \end{tikzpicture}
\\ PNM1   \hspace{6cm}PNM2 \\
\vspace{1cm}
  \begin{tikzpicture}
   \fill[gray!5, pattern=north west lines](0,5.5) rectangle (6,6);  
   \fill[gray!5, pattern=north west lines](0,0) rectangle (6,0.5);  
   \fill[gray!5, pattern=north west lines](0,.5) rectangle (0.5,5.5);  
   \fill[gray!5, pattern=north west lines](5.5,.5) rectangle (6,5.5);  
   \draw [step=.5, very thin, lightgray] (0,0) grid (6,6);
     \fill[gray!60](0+0.5,0.5+0.5) rectangle (0.5+0.5,1+0.5);  
     \fill[gray!60](1+0.5,0.5+0.5) rectangle (1.5+0.5,1+0.5);  
     \fill[gray!60](2+0.5,0.5+0.5) rectangle (2.5+0.5,1+0.5);  
     \fill[gray!60](3+0.5,0.5+0.5) rectangle (3.5+0.5,1+0.5);  
     \fill[gray!60](4+0.5,0.5+0.5) rectangle (4.5+0.5,1+0.5);  
     \fill[gray!60](0+0.5,1.5+0.5) rectangle (0.5+0.5,2+0.5);  
     \fill[gray!60](1+0.5,1.5+0.5) rectangle (1.5+0.5,2+0.5);  
     \fill[gray!60](2+0.5,1.5+0.5) rectangle (2.5+0.5,2+0.5);  
     \fill[gray!60](3+0.5,1.5+0.5) rectangle (3.5+0.5,2+0.5);  
     \fill[gray!60](4+0.5,1.5+0.5) rectangle (4.5+0.5,2+0.5);  
     \fill[gray!60](0+0.5,2.5+0.5) rectangle (0.5+0.5,3+0.5);  
     \fill[gray!60](1+0.5,2.5+0.5) rectangle (1.5+0.5,3+0.5);  
     \fill[gray!60](2+0.5,2.5+0.5) rectangle (2.5+0.5,3+0.5);  
     \fill[gray!60](3+0.5,2.5+0.5) rectangle (3.5+0.5,3+0.5);  
     \fill[gray!60](4+0.5,2.5+0.5) rectangle (4.5+0.5,3+0.5);  
     \fill[gray!60](0+0.5,3.5+0.5) rectangle (0.5+0.5,4+0.5);  
     \fill[gray!60](1+0.5,3.5+0.5) rectangle (1.5+0.5,4+0.5);  
     \fill[gray!60](2+0.5,3.5+0.5) rectangle (2.5+0.5,4+0.5);  
     \fill[gray!60](3+0.5,3.5+0.5) rectangle (3.5+0.5,4+0.5);  
     \fill[gray!60](4+0.5,3.5+0.5) rectangle (4.5+0.5,4+0.5);  
     \fill[gray!60](0+0.5,4.5+0.5) rectangle (0.5+0.5,5+0.5);  
     \fill[gray!60](1+0.5,4.5+0.5) rectangle (1.5+0.5,5+0.5);  
     \fill[gray!60](2+0.5,4.5+0.5) rectangle (2.5+0.5,5+0.5);  
     \fill[gray!60](3+0.5,4.5+0.5) rectangle (3.5+0.5,5+0.5);  
     \fill[gray!60](4+0.5,4.5+0.5) rectangle (4.5+0.5,5+0.5);  
   \end{tikzpicture}
\hspace{1cm}
  \begin{tikzpicture}
   \fill[gray!5, pattern=north west lines](0,5.5) rectangle (6,6);  
   \fill[gray!5, pattern=north west lines](0,0) rectangle (6,0.5);  
   \fill[gray!5, pattern=north west lines](0,.5) rectangle (0.5,5.5);  
   \fill[gray!5, pattern=north west lines](5.5,.5) rectangle (6,5.5);  
   \draw [step=.5, very thin, lightgray] (0,0) grid (6,6);
     \fill[gray!60](0+0.5,0.5+0.5) rectangle (0.5+0.5,1+0.5);  
     \fill[gray!60](1+0.5,0.5+0.5) rectangle (1.5+0.5,1+0.5);  
     \fill[gray!60](2+0.5,0.5+0.5) rectangle (2.5+0.5,1+0.5);  
     \fill[gray!60](3+0.5,0.5+0.5) rectangle (3.5+0.5,1+0.5);  
     \fill[gray!60](4+0.5,0.5+0.5) rectangle (4.5+0.5,1+0.5);  
     \fill[gray!60](0+0.5,1.5+0.5) rectangle (0.5+0.5,2+0.5);  
     \fill[gray!60](1+0.5,1.5+0.5) rectangle (1.5+0.5,2+0.5);  
     \fill[gray!60](2+0.5,1.5+0.5) rectangle (2.5+0.5,2+0.5);  
     \fill[gray!60](3+0.5,1.5+0.5) rectangle (3.5+0.5,2+0.5);  
     \fill[gray!60](4+0.5,1.5+0.5) rectangle (4.5+0.5,2+0.5);  
     \fill[gray!60](0+0.5,2.5+0.5) rectangle (0.5+0.5,3+0.5);  
     \fill[gray!60](1+0.5,2.5+0.5) rectangle (1.5+0.5,3+0.5);  
     \fill[gray!60](2+0.5,2.5+0.5) rectangle (2.5+0.5,3+0.5);  
     \fill[gray!60](3+0.5,2.5+0.5) rectangle (3.5+0.5,3+0.5);  
     \fill[gray!60](4+0.5,2.5+0.5) rectangle (4.5+0.5,3+0.5);  
     \fill[gray!60](0+0.5,3.5+0.5) rectangle (0.5+0.5,4+0.5);  
     \fill[gray!60](1+0.5,3.5+0.5) rectangle (1.5+0.5,4+0.5);  
     \fill[gray!60](2+0.5,3.5+0.5) rectangle (2.5+0.5,4+0.5);  
     \fill[gray!60](3+0.5,3.5+0.5) rectangle (3.5+0.5,4+0.5);  
     \fill[gray!60](4+0.5,3.5+0.5) rectangle (4.5+0.5,4+0.5);  
     \fill[gray!60](0+0.5,4.5+0.5) rectangle (0.5+0.5,5+0.5);  
     \fill[gray!60](1+0.5,4.5+0.5) rectangle (1.5+0.5,5+0.5);  
     \fill[gray!60](2+0.5,4.5+0.5) rectangle (2.5+0.5,5+0.5);  
     \fill[gray!60](3+0.5,4.5+0.5) rectangle (3.5+0.5,5+0.5);  
     \fill[gray!60](4+0.5,4.5+0.5) rectangle (4.5+0.5,5+0.5);  
     \fill[gray!60](0.5+0.5,0+0.5) rectangle (1+0.5,0.5+0.5);  
     \fill[gray!60](1.5+0.5,0+0.5) rectangle (2+0.5,0.5+0.5);  
     \fill[gray!60](2.5+0.5,0+0.5) rectangle (3+0.5,0.5+0.5);  
     \fill[gray!60](3.5+0.5,0+0.5) rectangle (4+0.5,0.5+0.5);  
     \fill[gray!60](4.5+0.5,0+0.5) rectangle (5+0.5,0.5+0.5);  
     \fill[gray!60](0.5+0.5,1+0.5) rectangle (1+0.5,1.5+0.5);  
     \fill[gray!60](1.5+0.5,1+0.5) rectangle (2+0.5,1.5+0.5);  
     \fill[gray!60](2.5+0.5,1+0.5) rectangle (3+0.5,1.5+0.5);  
     \fill[gray!60](3.5+0.5,1+0.5) rectangle (4+0.5,1.5+0.5);  
     \fill[gray!60](4.5+0.5,1+0.5) rectangle (5+0.5,1.5+0.5);  
     \fill[gray!60](0.5+0.5,2+0.5) rectangle (1+0.5,2.5+0.5);  
     \fill[gray!60](1.5+0.5,2+0.5) rectangle (2+0.5,2.5+0.5);  
     \fill[gray!60](2.5+0.5,2+0.5) rectangle (3+0.5,2.5+0.5);  
     \fill[gray!60](3.5+0.5,2+0.5) rectangle (4+0.5,2.5+0.5);  
     \fill[gray!60](4.5+0.5,2+0.5) rectangle (5+0.5,2.5+0.5);  
     \fill[gray!60](0.5+0.5,3+0.5) rectangle (1+0.5,3.5+0.5);  
     \fill[gray!60](1.5+0.5,3+0.5) rectangle (2+0.5,3.5+0.5);  
     \fill[gray!60](2.5+0.5,3+0.5) rectangle (3+0.5,3.5+0.5);  
     \fill[gray!60](3.5+0.5,3+0.5) rectangle (4+0.5,3.5+0.5);  
     \fill[gray!60](4.5+0.5,3+0.5) rectangle (5+0.5,3.5+0.5);  
     \fill[gray!60](0.5+0.5,4+0.5) rectangle (1+0.5,4.5+0.5);  
     \fill[gray!60](1.5+0.5,4+0.5) rectangle (2+0.5,4.5+0.5);  
     \fill[gray!60](2.5+0.5,4+0.5) rectangle (3+0.5,4.5+0.5);  
     \fill[gray!60](3.5+0.5,4+0.5) rectangle (4+0.5,4.5+0.5);  
     \fill[gray!60](4.5+0.5,4+0.5) rectangle (5+0.5,4.5+0.5);  
 \end{tikzpicture}
\\ PNM3   \hspace{6cm} PNM4 
\end{center}
\caption{{\bf The four spatial densities considered for the  PNM.} Gray squares represent monitored fields.
Dashed squares represent side fields where no treatment (and consequently, no decision) was applied.}
\label{fig:PMNs}
\end{figure}

\newpage
\subsubsection*{Three initial positions of infected fields for simulation}
\begin{figure}[!h]
\begin{center}
 \begin{tikzpicture}
   \fill[gray!5, pattern=north west lines](0,5.5) rectangle (6,6);  
   \fill[gray!5, pattern=north west lines](0,0) rectangle (6,0.5);  
   \fill[gray!5, pattern=north west lines](0,.5) rectangle (0.5,5.5);  
   \fill[gray!5, pattern=north west lines](5.5,.5) rectangle (6,5.5);  
   \fill[red!20](0.5,4.5) rectangle (1.5,5.5); 
   \fill[red!20](2.5,4.5) rectangle (3.5,5.5);
   \fill[red!20](2,3) rectangle (3,4); 
   \draw [step=.5, very thin, lightgray] (0,0) grid (6,6); 
  \end{tikzpicture}   
\end{center}
\caption{{\bf Position of infected fields at the beginning of simulations}. Simulations start with four grouped infected fields (red squares). Three different positions were used corresponding to the 3 red 2 x 2 squares.
\label{fig:initial-states}}
\end{figure}
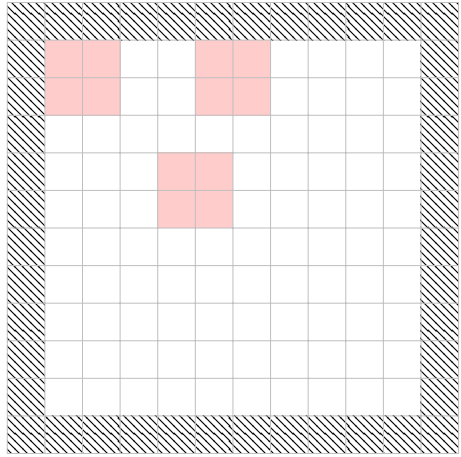

\subsubsection*{Sensitivity analysis}


We recall that the sensitivity analysis was performed for PMN3 and $h=1$, and that the factors are $\epsilon, \rho, \nu, c_{pest}, \gamma, q$. Settings for a simulation remain the same as for the comparison of the different PMNs.

\begin{figure}[!h]
  \includegraphics[width=13cm]{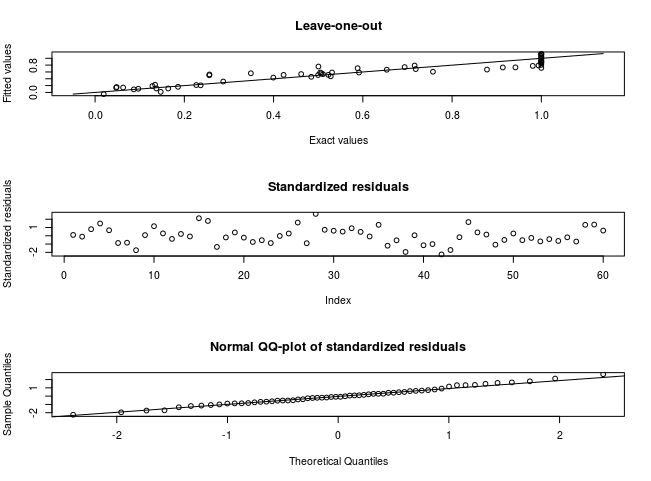}
  \caption{\textbf{Evaluation of Kriging model with constant trend for criterion T.}}
  \label{fig:T_m1}
\end{figure}

\begin{figure}[!h]
  \includegraphics[width=13cm]{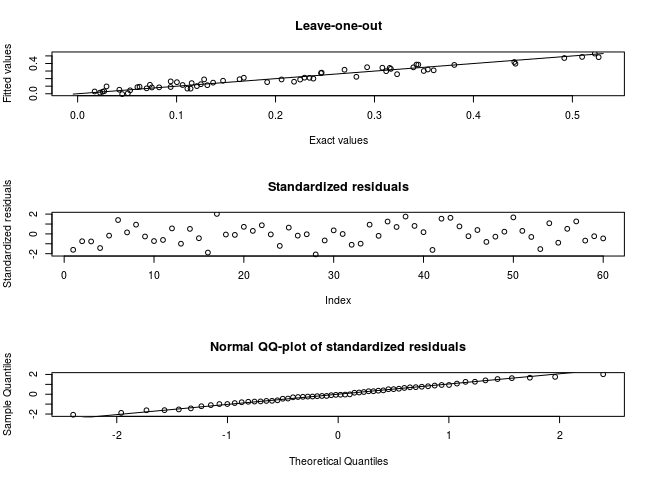}
  \caption{\textbf{Evaluation of Kriging model with constant trend for criterion I.}}
  \label{fig:I_m1}
\end{figure}

\begin{figure}[!h]
  \includegraphics[width=13cm]{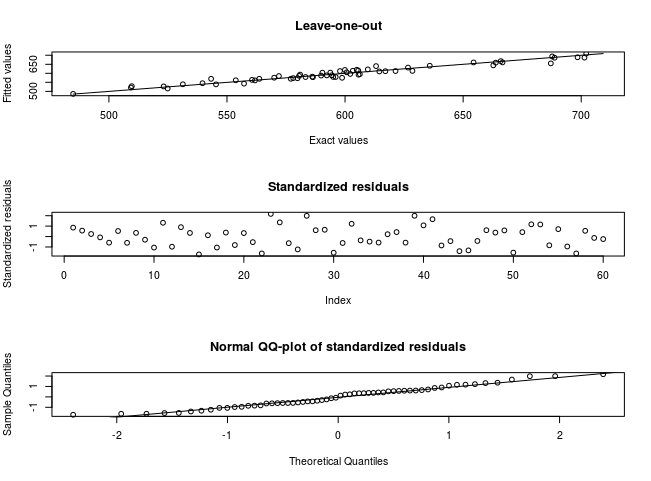}
  \caption{\textbf{Evaluation of Kriging model with constant trend for criterion R.}}
  \label{fig:r_m1}
\end{figure}

Computing mean values of $I$, $T$ and $R$ directly from the DBN model of pest dynamics combined with the decision rule requires running many simulations. 
Therefore, in order to perform the sensitivity analysis, we first created a metamodel to  approximate the relationship between  the 6 factors and  mean values of the three criteria.
The metamodel is estimated from samples of the simulator. We chose 60  values for the vector $(\epsilon, \rho, \nu, c_{pest}, \gamma, q)$  according to a Latin Hypercube Sampling, obtained by combining 10 values for each factor, uniformly sampled in the domain of variation of each factor.
We then fitted a Kriging model with constant trend on these 60 samples (function km of R package DiceKriging). We did not consider a more complex Kriging model since this one   
was well adapted for each criterion according to classical evaluation methods (see figures ~\ref{fig:T_m1}-\ref{fig:r_m1}).

To evaluate the Sobol indices, we performed a Kriging-based global sensitivity analysis taking  both the error from using a metamodel and the error from estimating the Sobol indices by Monte-Carlo into account. The three points corresponding to the three pests teste are very distant in the  hypercube of the factor domains,  so we considered that factors values too far from these points may not be representative of any existing pest dynamics. We therefore estimated distinct indices for each pest type, by  reducing the domains of variation of each factor to domains centered around the expert value for the given pest used to perform the PMN comparison (see Table 2 in the article).

\end{document}